\documentclass{aastex63}
\usepackage{upgreek}
\usepackage{textcomp}
\usepackage{booktabs}
\usepackage{ragged2e}
\usepackage{rotating}
\usepackage{sidecap}
\usepackage{multirow}
\received{}
\revised{}
\accepted{}
\submitjournal{ApJ}

\shorttitle{Radial transport across Jupiter}
\shortauthors{van Kooten et al.}

\begin{document}

\title{Hybrid accretion of carbonaceous chondrites by radial transport across the Jupiter barrier}

\correspondingauthor{Elishevah van Kooten}
\email{elishevah.vankooten@sund.ku.dk}

\author[0000-0003-3333-4421]{Elishevah van Kooten}
\affiliation{Centre for Star and Planet Formation and Globe Institute, University of Copenhagen \\
DK-1350 Copenhagen, Denmark}
\affiliation{Universit\'{e} de Paris, Institut de Physique du Globe de Paris\\ CNRS UMR 7154, 1 rue Jussieu, 75238 Paris, France}

\author[0000-0003-4149-0627]{Martin Schiller}
\affiliation{Centre for Star and Planet Formation and Globe Institute, University of Copenhagen \\\O{}ster Voldgade 5-7, DK-1350 Copenhagen, Denmark}

\author{Fr\'{e}d\'{e}ric Moynier}
\affiliation{Universit\'{e} de Paris, Institut de Physique du Globe de Paris\\ CNRS UMR 7154, 1 rue Jussieu, 75238 Paris, France}

\author{Anders Johansen}
\affiliation{Lund Observatory, Department of Astronomy and Theoretical Physics, Lund University \\ S\"{o}lvegatan 27, SE-221 00 Lund, Sweden}
\affiliation{Centre for Star and Planet Formation and Globe Institute, University of Copenhagen \\
DK-1350 Copenhagen, Denmark}

\author{Troels Haugb\o{}lle}
\affiliation{Centre for Star and Planet Formation and Globe Institute, University of Copenhagen \\
DK-1350 Copenhagen, Denmark}
\affiliation{Niels Bohr Institute, University of Copenhagen \\
DK-2100 Copenhagen, Denmark}

\author[0000-0001-9966-2124]{Martin Bizzarro}
\affiliation{Centre for Star and Planet Formation and Globe Institute, University of Copenhagen \\
DK-1350 Copenhagen, Denmark}
\affiliation{Universit\'{e} de Paris, Institut de Physique du Globe de Paris\\ CNRS UMR 7154, 1 rue Jussieu, 75238 Paris, France}

\begin{abstract}
Understanding the origin of chondritic components and their accretion pathways is critical to unravel the magnitude of mass transport in the protoplanetary disk, the accretionary history of the terrestrial planet region and, by extension, its prebiotic inventory. Here, we trace the heritage of pristine components from the relatively unaltered CV chondrite Leoville through their mass-independent Cr and mass-dependent Zn isotope compositions. Investigating these chondritic fractions in such detail reveals an onion-shell structure of chondrules, which is characterized by $^{54}$Cr- and $^{66}$Zn-poor cores surrounded by increasingly $^{54}$Cr- and $^{66}$Zn-rich igneous rims and an outer coating of fine-grained dust. This is interpreted as a progressive addition of $^{54}$Cr- and $^{66}$Zn-rich, CI-like material to the accretion region of these carbonaceous chondrites. Our findings show that the observed Cr isotopic range in chondrules from more altered CV chondrites is the result of chemical equilibration between chondrules and matrix during secondary alteration. The $^{54}$Cr-poor nature of the cores of Leoville chondrules implies formation in the inner Solar System and subsequent massive outward chondrule transport past the Jupiter barrier. At the same time, CI-like dust is transferred inwards. We propose that the accreting Earth acquired CI-like dust through this mechanism within the lifetime of the disk. This radial mixing of chondrules and matrix shows the limited capacity of Jupiter to act as an efficient barrier and maintain the proposed non-carbonaceous and carbonaceous chondrite dichotomy over time. Finally, also considering current astrophysical models, we explore both inner and outer Solar System origins for the CV chondrite parent body. 

\end{abstract}


\section{Introduction} \label{sec:intro}
Chondrites are fragments of primitive planetesimals and represent sedimentary agglomerates comprising the Solar System’s oldest and least altered planetary building blocks. They primarily consist of three components, namely chondrules, matrix and less abundant refractory inclusions, which represent the first formed solids in the protoplanetary disk \citep{connelly_absolute_2012}. Chondrules are spherical objects believed to have formed from melting of free-floating dust agglomerates during transient heating events in the disk. The matrix present in chondrites is thought to consist of relatively fine-grained, unaltered material that surrounds the other chondrite components. Collectively, chondrites and their components provide a unique time window into the early evolution of the gaseous protoplanetary disk when the planets accreted their main masses. These objects are typically divided into non-carbonaceous (NC) and carbonaceous (CC) chondrites, and it has been suggested that the former accreted in the terrestrial planet region (i.e., sunward of Jupiter), whereas the latter accreted in the accretion region of gas giant planets. This divergent heritage of chondrites is based on the relatively volatile-rich nature of CC, the distribution of chondritic components in the chondrite groups and their distinct isotope systematics relative to NC \citep{budde_molybdenum_2016,warren_stable-isotopic_2011}. \\

The proposed early establishment of such a dichotomy through the rapid accretion of Jupiter \citep{kruijer_age_2017} implies limited or no chemical and/or isotopic exchange between the NC and CC reservoirs. This model is in line with the proposed elemental and isotopic complementarity between chondrules and matrix, which suggests that chondrite components are genetically linked and formed from a common disk reservoir. Both the complementarity \citep{hezel_chemical_2010} and the dichotomy model do not allow for mass transport in the protoplanetary disk, but view planetary accretion within a more static environment, where planetary building blocks are sampled within the direct vicinity of the accreting object. The assumed absence of significant mass transport between the inner terrestrial planet accretion region and the outer Solar System is inconsistent with recent isotopic evidence indicating protracted accretion of volatile-rich matter to the terrestrial planets during the lifetime of the protoplanetary disk \citep{schiller_iron_2020,schiller_isotopic_2018}. In detail, the nucleosynthetic Ca and Fe isotope composition of the Earth requires mixing between an inner Solar System derived endmember and a CI-like endmember from the outer Solar System. CI (Ivuna-type) chondrites are volatile-rich (i.e., rich in water and organics) meteorites closely matching the composition of the solar photosphere and consist mainly of fine-grained dust (i.e., matrix). Although it has been postulated that a sharp Mo isotopic dichotomy exists between inner and outer Solar System objects \citep{budde_molybdenum_2016}, the observation that Earth falls in between NC and CC trends \citep{budde_molybdenum_2019} appears to contradict the view that these two reservoirs existed in the inner and outer Solar System, respectively, and maintained isolation from each other. This observation has been interpreted as evidence for an outer Solar System origin for the Moon-forming impactor, which was admixed to Earth’s mantle during a giant impact. However, this hypothesis is unlikely since the Moon and Earth are isotopically nearly identical for most elements \citep{mastrobuono-battisti_primordial_2015,zhang_proto-earth_2012}. Alternatively, the Mo isotope composition of Earth may reflect progressive admixing of volatile-rich matter to the inner Solar System as inferred from the terrestrial Fe and Ca nucleosynthetic compositions. A better understanding of the extent of mass transport between the NC and CC reservoirs is critical to elucidate Earth’s accretion history, including the nature of volatiles and prebiotic molecules delivered to our planet. For example, in a model of limited mass transport between the inner and outer Solar System, delivery of abundant prebiotic molecules such as complex organic compounds critical to life occurs by stochastic processes late in the history of the terrestrial planets. In contrast, progressive inward mass transport of outer Solar System material during the disk lifetime raises the possibility that the delivery of prebiotic molecules to the terrestrial planets is a natural consequence of their main accretion phase, rather than a stochastic coincidence.\\

Recent astronomical observations of young protoplanetary disks and planet formation simulations suggest that the growth of rocky planets occurs by the highly efficient mechanism of pebble accretion, that is, the accretion of millimeter- to centimeter-sized particles onto planetesimal seeds during the $\sim$5 Myr protoplanetary disk lifetime \citep{johansen_growth_2015,johansen_pebble_2021}. In the inner Solar System, this process is believed to be driven by the accretion of mm-sized chondrules. Both the petrology and individual Pb-Pb ages of chondrules \citep{bollard_early_2017} indicate that these objects experienced multiple melting events. For example, igneous rims surrounding chondrule cores in CV (Vigarano-type) chondrites record the accretion of additional dust to the chondrules followed by a heating event that likely post-dates the melting of the corresponding cores \citep{rubin_chondrules_1987,van_kooten_zinc_2019}. Moreover, chondrules are often characterized by fine-grained dust rims (FGRs), which surround the igneous rims and represent an addition of even more pristine dust to the chondrule system. Thus, individual chondrules are time-sequenced samples that can provide insights into the secular composition and isotopic evolution of disk solids that accreted to planets. In particular, the onion-shell structure of a chondrule and its rims can be regarded as a miniature planet and its feeding zones. Thus, by sampling the individual accretion layers of the chondrule system, it is possible to reconstruct the flux of material in the early Solar System during the time when the terrestrial planets acquired their main mass.\\

Here, we have sampled chondrule cores and their igneous rims, as well as their fine-grained dust rims and more coarse-grained intra-chondrule matrix (ICM) from the relatively unaltered CV3.1 chondrite Leoville. The aim is to investigate their genetic heritage and volatile related processes through mass-independent Cr isotope and mass-dependent Zn isotope systematics, respectively. In detail, nucleosynthetic Cr isotope signatures have been used on bulk chondrites and chondrules in the past to trace the origin of their precursor materials within the protoplanetary disk \citep{olsen_magnesium_2016,trinquier_widespread_2007,van_kooten_isotopic_2016,warren_stable-isotopic_2011}, whereas Zn isotopes have been used as a probe for evaporation and condensation processes of chondrites and chondrules \citep{luck_zn_2005,pringle_origin_2017,van_kooten_zinc_2019}. Hence, combining these isotope systems will provide us with useful information regarding the formation location and corresponding thermodynamic conditions of various chondritic components as well as the degree and direction of their inferred mass transport. CV chondrites have been chosen for this study because of the large size of their components relative to other carbonaceous chondrites, which allows for detailed sampling. We show that from such a detailed investigation of one of the least altered CV chondrites, a completely new picture emerges for the accretion history of the CV chondrite parent body that impacts our understanding of the accretion history of terrestrial planets.

\section{Results} \label{sec:results}
\subsection{The petrology and compositions of Leoville fractions} \label{sec:petrology}
\begin{figure}[ht!]
\plotone{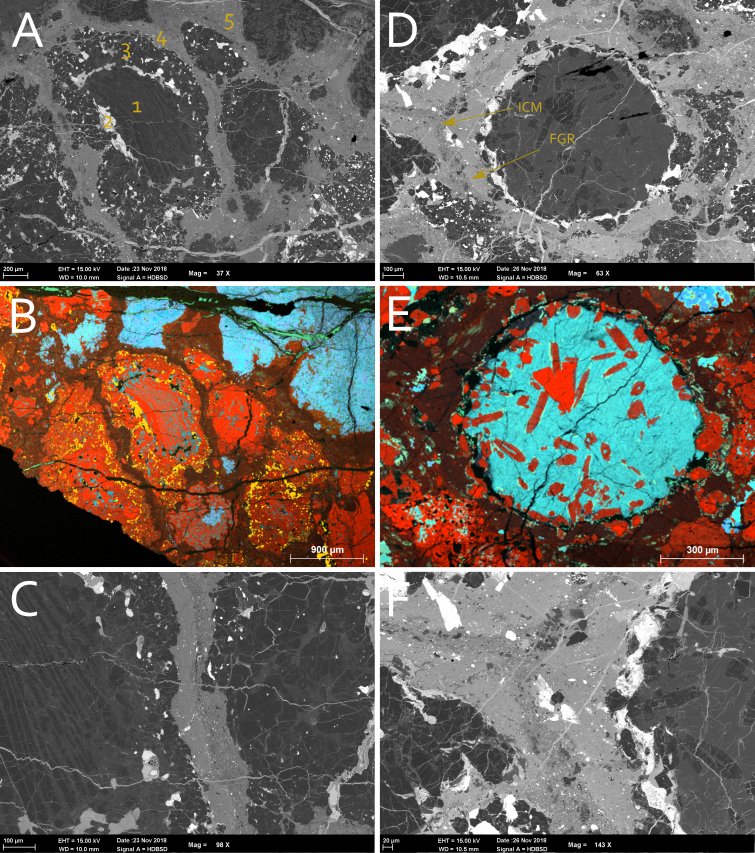}
\caption{Back scattered electron images and Mg-Ca-Al-S (red-green-blue-yellow) elemental maps from Leoville (CV3.1) chondrules and matrix. A) depicting various components of Leoville, including 1) chondrule core with a barred olivine texture (sample Ach4), 2) metal/sulfide rim around the core, 3) igneous rim with forsterite, low-Ca pyroxene and abundant metal/sulfide grains, 4) a fine-grained dust rim and 5) the intra-chondrule matrix. B) Elemental map of A), where yellow reflects the sulfides. The bright blue grain in the upper right corner is a CAI. C) Close-up of A), in which the distinction between a FGR and ICM is highlighted. D) An Al-rich chondrule (sample Bch6) with a single olivine grain and pyroxenes, surrounded by a metal rim and a FGR. The distinction between FGR and ICM is highlighted. E) Elemental map of D). F) Close-up of D), which highlights the borders between three FGRs and the ICM. For more images of Leoville components sampled in this study, see the Appendix B. \label{fig:petro1}}
\end{figure}

\begin{figure}[ht!]
\plotone{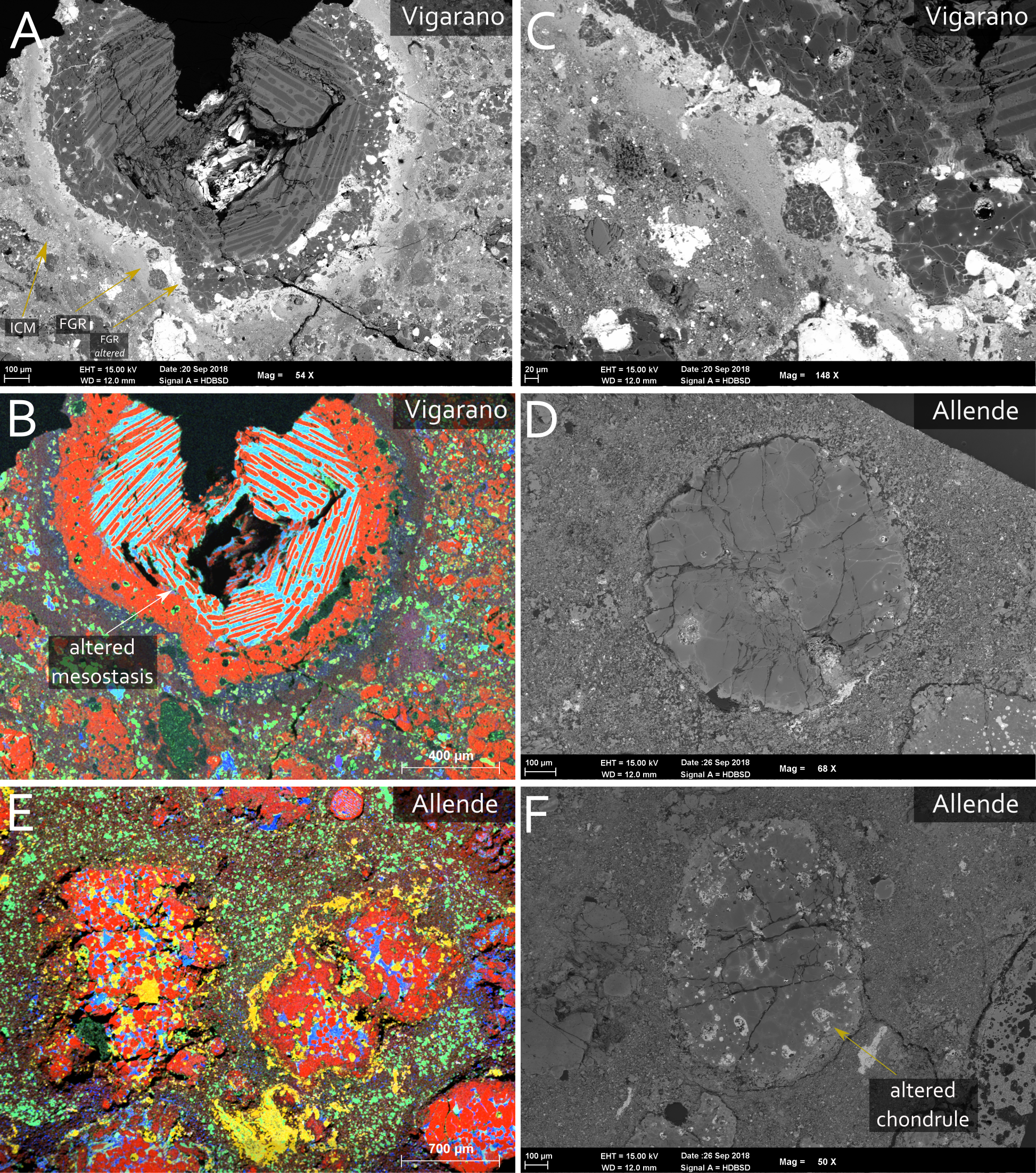}
\caption{Back scattered electron images and Mg-Ca-Al-S (red-green-blue-yellow) elemental maps from Vigarano (CV3.1-3.4) and Allende (CV$>$3.6) chondrules and matrix. Vigarano contains more and less altered areas. A) shows a more altered area with a barred type I chondrule with mineralogical zonation. The FGR surrounding the chondrule is altered near the chondrule boundary where exchange of among others Fe-Mg occurs. B) Elemental map of A). C) Close-up of A). D) Chondrule from Allende. In the surrounding coarse-grained matrix, the distinction between FGR and ICM cannot be made because of secondary alteration. E) Elemental map of Allende chondrules, where the alteration of the chondrules is reflected by the transfer of sulfides to the chondrule cores and the alteration of the coarse-grained matrix is visible. The green grains in the matrix are recrystallized Ca-pyroxenes. F) Another chondrule from Allende where the rim of the chondrule is significantly altered through Fe-Mg elemental exchange.  
\label{fig:petro2}}
\end{figure}

We have investigated the petrology and elemental composition of Leoville fractions using high resolution back scattered electron images, elemental maps, laser ablation and inductively coupled plasma mass spectrometry (ICPMS) analyses (for a detailed description see Appendix B and Table A1 and A2) in this and previous studies \citep{van_kooten_unifying_2019,van_kooten_zinc_2019}. Our data demonstrate that Leoville, classified as a reduced CV3.1 chondrite \citep{bonal_thermal_2016}, is one of the least altered CV chondrites (Fig. \ref{fig:petro1} versus Fig. \ref{fig:petro2}) when compared to other CV chondrites such as Vigarano (CV3.1-3.4)  and Allende (CV3.6). Here, we focus on chondrules with simple core-igneous rim pairs (Fig. \ref{fig:petro1}, Fig. \ref{fig:petrology}). All chondrules are surrounded by FGRs and the space between the FGRs is occupied by ICM, which includes chondrule fragments and larger sulfide and metal grains. We note that for more altered CV chondrites such as Vigarano and Allende, these two types of matrix become indistinguishable (Fig. \ref{fig:petro2}). The fluid-assisted thermal metamorphism results in a coarsening and dehydration of the entire matrix towards a homogeneous texture of phyllosilicates and larger grains of anhydrous Ca-rich pyroxene. This alteration begins with the leaching of siderophile elements from the oxidized/sulfurized metal into the matrix and modification of the chondrule mesostasis (Fig. \ref{fig:petro2}B) and concludes with the larger-scale elemental exchange between chondrules and matrix (Fig. \ref{fig:petro2}D and Fig. \ref{fig:petro2}F). Furthermore, we observe a transfer of a sulfur-rich matrix to the production of sulfide grains in the chondrules (Fig. \ref{fig:petro2}E, \citealp{van_kooten_zinc_2019}). The pristine Leoville chondrite overall lacks these alteration features and, thus, by focusing on this chondrite, we can be confident that the chondrules and their rims record primary chemical and isotopic information. 

\subsection{Cr isotope signatures of Leoville components} \label{sec:Cr_isotopes}
\begin{figure}[ht!]
\centering
\includegraphics[width=0.55\textwidth]{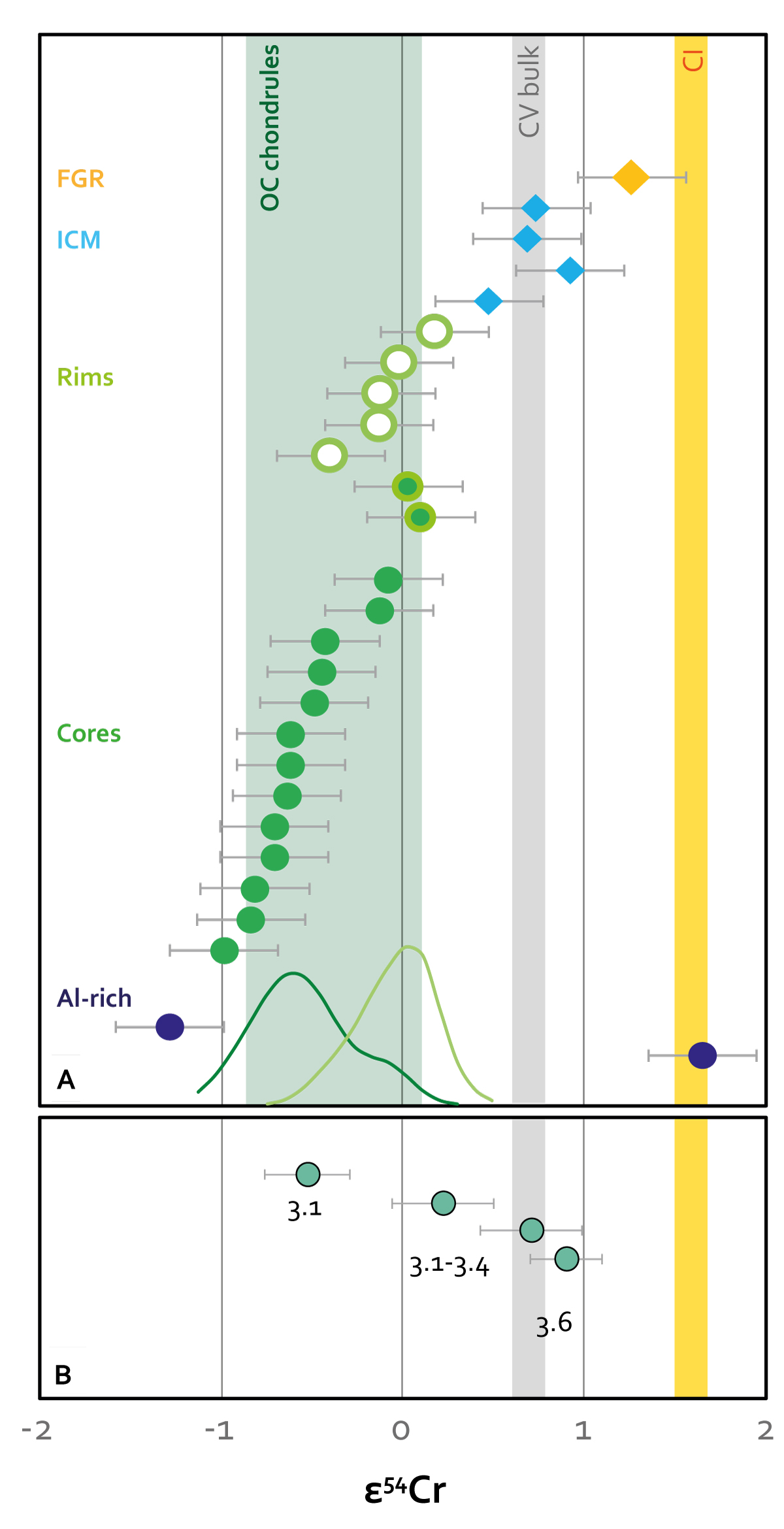}
\caption{$\upepsilon$$^{54}$Cr values from Leoville components (A) Data from Table \ref{tab:CrZn}, where the dark blue circles reflect the Al-rich chondrules, solid green shows the Leoville chondrule cores, solid green with light green rim are the two bulk chondrules sampled (including core and rim), open green circles are the chondrule rims, blue diamonds reflect the clean ICM and the yellow diamond represents the FGR. We also show the range of OC chondrules (shaded green area, \citealp{bollard_combined_2019}), the CV bulk $\upepsilon$$^{54}$Cr value and the $\upepsilon$$^{54}$Cr value for CI chondrites. All errors reflect the 2SD external reproducibility of the measurements. A density distribution is shown at the bottom for the chondrule cores and rims. B) $\upepsilon$$^{54}$Cr weighted averages for chondrule populations from Leoville (CV3.1), Vigarano (CV3.1-3.4, \citealp{olsen_magnesium_2016}), NWA 3118 (CV3.6, \citealp{olsen_magnesium_2016}) and Allende (CV$>$3.6,  \citealp{kadlag_cr_2019}). 
\label{fig:e54Cr}}
\end{figure}

We have measured the Cr isotope compositions of 11 ICM fragments as well as one FGR (Appendix A for detailed methods, Table \ref{tab:CrZn}, Fig. \ref{fig:e54Cr}), which have also been analyzed for their Zn isotope compositions on the same aliquots. Given that the average thickness of the FGRs is $\sim$100 $\upmu$m, which is comparable to that of the microdrill bits, it was not possible to sample additional FGRs without contamination from surrounding materials. However, we note that the chemical composition of the FGRs in Leoville and in similarly unaltered sections of Vigarano are identical and, as such, the obtained isotope compositions of the FGRs are likely representative of CV chondrites in general \citep{van_kooten_unifying_2019}. All errors reported here are 2SD. We distinguish between clean and contaminated (chondrule fragments present) ICM (see Appendix B). We show that the average $\upepsilon$$^{54}$Cr value of the clean ICM (0.70$\pm$0.37 ‱) is indistinguishable from that of our bulk CV chondrite value of 0.81$\pm$0.28 ‱ and averaged literature values of 0.86$\pm$0.06 ‱. The $^{55}$Mn/$^{52}$Cr ratios (0.59$\pm$0.24) and $\upepsilon$$^{53}$Cr ($^{53}$Cr being the decay product of $^{53}$Mn with a half-life of 3.74 Myr) signatures (0.20$\pm$0.51 ‱) of the clean ICM are also consistent with a bulk CV chondrite signature. The contaminated ICM is more variable and ranges between –2.20$\pm$0.35 ‱ and 0.51$\pm$0.23 ‱, with most $\upepsilon$$^{54}$Cr values being negative (Table \ref{tab:CrZn}). The FGR has a $\upepsilon$$^{54}$Cr value of 1.26$\pm$0.32 ‱, which is within error of the CI chondrite value of 1.54$\pm$0.30 ‱ measured in this study and literature values \citep{qin_contributors_2010,schiller_precise_2014,trinquier_widespread_2007}. Moreover, its $\upepsilon$$^{53}$Cr signature (0.24$\pm$0.12 ‱) and $^{55}$Mn/$^{52}$Cr ratio of 0.86 also correspond to CI chondrites \citep{qin_contributors_2010,shukolyukov_manganesechromium_2006,van_kooten_role_2020}. The type I chondrule cores have an average $\upepsilon$$^{54}$Cr of –0.58$\pm$0.16 ‱ (n = 13, 2SD), with a range between –1.00$\pm$0.30 ‱ and –0.09$\pm$0.30 ‱. This overlaps with the range of previous individual CV chondrule Cr isotope analyses ($\upepsilon$$^{54}$Cr = –0.79$\pm$0.10 to 2.01$\pm$0.10 ‱; \citealp{olsen_magnesium_2016}), but the distribution of the $\upepsilon$$^{54}$Cr values and the average obtained for the Leoville chondrules is not in agreement with previous datasets \citep{kadlag_cr_2019,olsen_magnesium_2016,williams_chondrules_2020}. In detail, the average $\upepsilon$$^{54}$Cr values of the Leoville chondrule cores is significantly more negative than for Vigarano chondrules ($\upepsilon$$^{54}$Cr = 0.26$\pm$0.28 ‱, n = 10; \citealp{olsen_magnesium_2016}), NWA 3118 ($\upepsilon$$^{54}$Cr = 0.77$\pm$0.28 ‱, n = 6; \citealp{olsen_magnesium_2016}) and bulk Allende chondrules ($\upepsilon$$^{54}$Cr = 0.90$\pm$0.28 ‱, n = 100; \citealp{kadlag_cr_2019}). The $\upepsilon$$^{53}$Cr signatures of the chondrule cores are –0.20$\pm$0.09 ‱, within error of the solar system initial $\upepsilon$$^{53}$Cr value of –0.18 ‱ \citep{gopel_mn-cr_2015}, and the cores have an average $^{55}$Mn/$^{52}$Cr ratio of 0.20$\pm$0.14. The five igneous rims have higher $\upepsilon$$^{54}$Cr values with an average of 0.10$\pm$0.24 ‱ and ranging between –0.46$\pm$0.30 and 0.22$\pm$0.30 ‱. The $\upepsilon$$^{54}$Cr value of each individual chondrule rim is higher than that of the corresponding core (Table \ref{tab:CrZn}). The $\upepsilon$$^{53}$Cr signatures of rims vary little and average –0.32$\pm$0.11 ‱  despite exhibiting variable $^{55}$Mn/$^{52}$Cr ratios between 0.10 and 0.77. Finally, we have measured two bulk chondrules (i.e., core with igneous rim) with $\upepsilon$$^{54}$Cr values of 0.02$\pm$0.40 ‱ and 0.09$\pm$0.18 ‱ and two Al-rich chondrules with $\upepsilon$$^{54}$Cr values of –1.30$\pm$0.11 ‱ and 1.63$\pm$0.30 ‱.

\begin{longtable}[ht!]{lcrm{0.25cm}p{1.5cm}rm{0.25cm}p{1.5cm}rm{0.25cm}p{1cm}}
\caption{Mass-independent Cr and mass-dependent Zn isotope data of Leoville components and reference materials. For Cr isotope analyses, in case of samples measured on multiple filaments, the error is taken as the 2SD of the weighted mean is taken, otherwise the error reflects the external reproducibility of the measurements taken from the standards (30 ppm [2SD] on $\upepsilon$$^{54}$Cr). WM = weighted mean. *Data from \citet{van_kooten_zinc_2019}. The row ‘aliquots’ for the standards reflects the number of aliquots processed through chemistry (Appendix A for detailed explanation). FGR = fine-grained rim and ICM = intra chondrule matrix. \label{tab:CrZn}}\\
\toprule
 &  & \multicolumn{3}{c}{\textbf{$\updelta$$^{66}$Zn}} & \multicolumn{3}{c}{\textbf{$\upepsilon$$^{54}$Cr}} & \multicolumn{3}{c}{\textbf{$\upepsilon$$^{53}$Cr}} \\
 \hline
\endfirsthead
\multicolumn{11}{c}%
	{\tablename\ \thetable\ -- \textit{Continued from previous page}} \\
\hline
&  & \multicolumn{3}{c}{\textbf{$\updelta$$^{66}$Zn}} & \multicolumn{3}{c}{\textbf{$\upepsilon$$^{54}$Cr}} & \multicolumn{3}{c}{\textbf{$\upepsilon$$^{53}$Cr}} \\
\hline
\endhead
\hline 
\multicolumn{11}{r}{\textit{Continued on next page}} \\
\endfoot
\bottomrule
\endlastfoot
 &  & \textbf{} & \textbf{} &  & \multicolumn{1}{c}{\textbf{}} & \textbf{} &  & \multicolumn{1}{c}{\textbf{}} & \textbf{} &  \\
\textit{\textbf{Reference materials}} & \textit{Aliquots} &  & \multicolumn{1}{l}{} &  &  & \multicolumn{1}{l}{} &  &  & \multicolumn{1}{l}{} &  \\
NWA 12523 (CV bulk) & \textit{3} & 0.27* & ± & 0.19 & 0.81 & ± & 0.28 & 0.04 & ± & 0.42 \\
Ivuna (CI bulk) & \textit{5} &  & \multicolumn{1}{l}{} &  & 1.50 & ± & 0.23 & 0.15 & ± & 0.06 \\
PCC1 & \textit{2} &  & \multicolumn{1}{l}{} &  & 0.12 & ± & 0.09 & -0.01 & ± & 0.08 \\
BHVO2 & \textit{1} & 0.34* & ± & 0.08 & -0.03 &  &  & 0.02 & \multicolumn{1}{l}{} &  \\
 &  &  & \multicolumn{1}{l}{} &  &  & \multicolumn{1}{l}{} &  &  & \multicolumn{1}{l}{} &  \\
 &  &  & \multicolumn{1}{l}{} &  &  & \multicolumn{1}{l}{} &  &  & \multicolumn{1}{l}{} &  \\
\textit{\textbf{Leoville chondrules}} & \textit{Fraction} &  & \multicolumn{1}{l}{} &  &  & \multicolumn{1}{l}{} &  &  & \multicolumn{1}{l}{} &  \\
Bch3 & Al-rich core &  & \multicolumn{1}{l}{} &  & 1.63 & ± & 0.30 & 0.25 & ± & 0.15 \\
Bch6 & Al-rich core & 0.10 & ± & 0.12 & -1.30 & ± & 0.11 & -0.08 & ± & 0.25 \\
 &  &  & \multicolumn{1}{l}{} &  &  &  &  &  &  &  \\
Ach1 & core & -0.94 & ± & 0.12 & -0.63 & ± & 0.49 & -0.11 & ± & 0.09 \\
Ach7 & core &  & \multicolumn{1}{l}{} &  & -0.83 & ± & 0.39 & -0.38 & ± & 0.32 \\
Ach8 & core &  & \multicolumn{1}{l}{} &  & -0.85 & ± & 0.30 & -0.36 & ± & 0.15 \\
Ach4 & core &  & \multicolumn{1}{l}{} &  & -0.65 & ± & 0.30 & -0.11 & ± & 0.15 \\
Ach9 & core &  & \multicolumn{1}{l}{} &  & -0.63 & ± & 0.30 & -0.03 & ± & 0.15 \\
 &  &  & \multicolumn{1}{l}{} &  &  &  &  &  &  &  \\
Bch7 & core &  & \multicolumn{1}{l}{} &  & -0.72 & ± & 0.30 & -0.31 & ± & 0.15 \\
Ach10 & core &  & \multicolumn{1}{l}{} &  & -0.44 & ± & 0.30 & -0.16 & ± & 0.15 \\
Bch1 & core &  & \multicolumn{1}{l}{} &  & -0.50 & ± & 0.30 & 0.02 & ± & 0.15 \\
Bch2 & core &  & \multicolumn{1}{l}{} &  & -0.14 & ± & 0.30 & -0.1 & ± & 0.15 \\
 &  &  & \multicolumn{1}{l}{} &  &  &  &  &  &  &  \\
C1 & core & -0.77 & ± & 0.19 & -0.09 & ± & 0.30 & -0.38 & ± & 0.15 \\
C2 & core & -0.77 & ± & 0.19 & -1.00 & ± & 0.30 & -0.42 & ± & 0.15 \\
C3 & core & -0.90 & ± & 0.19 & -0.46 & ± & 0.30 & -0.67 & ± & 0.15 \\
C5 & core & -0.72 & ± & 0.19 & -0.72 & ± & 0.30 & -0.33 & ± & 0.15 \\
C6 & core & -0.77 & ± & 0.19 & \multicolumn{1}{r}{} & \multicolumn{1}{r}{} &  & \multicolumn{1}{r}{} &  &  \\
 &  &  &  &  &  & \multicolumn{1}{l}{} &  &  &  &  \\
Ach2 & core+rim & 0.15 & ± & 0.12 & 0.09 & ± & 0.18 & -0.12 & ± & 0.38 \\
Ach3 & core+rim & -0.06 & ± & 0.12 & 0.02 & ± & 0.40 & -0.16 & ± & 0.20 \\
 &  &  &  &  &  & \multicolumn{1}{l}{} &  &  &  &  \\
C1 & rim & -0.14 & ± & 0.12 & 0.17 & ± & 0.30 & -0.28 & ± & 0.15 \\
C2 & rim & -0.11 & ± & 0.12 & -0.41 & ± & 0.30 & -0.36 & ± & 0.15 \\
C3 & rim & -0.04 & ± & 0.12 & -0.14 & ± & 0.30 & -0.26 & ± & 0.15 \\
C5 & rim & 0.23 & ± & 0.12 & -0.13 & ± & 0.30 & -0.33 & ± & 0.15 \\
C6 & rim & 0.01 & ± & 0.12 & -0.03 & ± & 0.30 & -0.39 & ± & 0.15 \\
 &  &  & \multicolumn{1}{l}{} &  &  & \multicolumn{1}{l}{} &  &  &  &  \\
\textit{\textbf{Leoville matrix}} &  &  & \multicolumn{1}{l}{} &  &  & \multicolumn{1}{l}{} &  &  &  &  \\
Ach1mx & IC mx & -0.02 & ± & 0.12 & 0.47 & ± & 0.28 & -0.17 & ± & 0.04 \\
Ach2mx & IC mx & -0.02 & ± & 0.12 & -0.51 & ± & 0.33 & -0.11 & ± & 0.09 \\
Ach3mx & IC mx & -0.26 & ± & 0.12 & 0.51 & ± & 0.23 & 0.15 & ± & 0.05 \\
Ach7mx & IC mx & -0.19 & ± & 0.12 & 0.92 & ± & 0.11 & 0.26 & ± & 0.11 \\
Bmx1 & ICmx & 0.21 & ± & 0.12 & -0.21 & ± & 0.40 & 0.14 & ± & 0.17 \\
Bmx2 & IC mx & 0.14 & ± & 0.12 & -0.17 & ± & 0.30 & -0.17 & ± & 0.15 \\
Bmx3 & IC mx & -0.43 & ± & 0.12 & -2.20 & ± & 0.39 & 0.12 & ± & 0.07 \\
Leo2ch1mx1 & IC mx & -0.09 & ± & 0.12 & 0.68 & ± & 0.30 & 0.28 & ± & 0.15 \\
Leo2ch1mx2 & IC mx & -0.09 & ± & 0.12 & 0.73 & ± & 0.30 & 0.42 & ± & 0.15 \\
Leo2ch2mx & IC mx & -0.60 & ± & 0.12 & -0.08 & ± & 0.30 & 0.59 & ± & 0.15 \\
Leo2ch3mx & IC mx & -0.29 & ± & 0.12 & -0.47 & ± & 0.30 & 0.16 & ± & 0.15 \\
Ch1mx & FGR & 0.14 & ± & 0.12 &  &  &  &  &  &  \\
Ch2mx & FGR & 0.24 & ± & 0.12 & 1.26 & ± & 0.32 & 0.24 & ± & 0.12 \\
 &  &  & \multicolumn{1}{l}{} &  &  &  &  &  &  &  \\
 &  &  & \multicolumn{1}{l}{} &  &  &  &  &  &  &  \\
cores (WM) &  & -0.84 & ± & 0.10 & -0.58 & ± & 0.16 & -0.20 & ± & 0.09 \\
rims (WM) &  & -0.01 & ± & 0.18 & -0.11 & ± & 0.26 & -0.32 & ± & 0.07 \\
FGR &  & 0.24 & ± & 0.12 & 1.26 & ± & 0.32 & 0.24 & ± & 0.12 \\
ICM (clean) &  & -0.10 & ± & 0.14 & 0.70 & ± & 0.37 & 0.20 & ± & 0.51\\
\end{longtable}

\subsection{The Zn isotope compositions of Leoville components} \label{sec:Zn_isotopes}
We have measured the Zn isotope compositions of 11 ICM fragments (see Appendix A for detailed methods). Previous measurements of Leoville matrix include two FGRs (Table \ref{tab:CrZn}, Fig. \ref{fig:massdependent}, \citealp{van_kooten_zinc_2019}). Both $\updelta$$^{66}$Zn values determined for FGRs are indistinguishable (0.14$\pm$0.12 ‰ and 0.24$\pm$0.12 ‰) and form an average of 0.19$\pm$0.14‰. ICM samples are generally lighter with values ranging between –0.43$\pm$0.12 ‰ and 0.14$\pm$0.12 ‰. Samples regarded as representing pure ICM return an average $\updelta$$^{66}$Zn composition of –0.10$\pm$0.14 ‰, whereas contamination results in more variable $\updelta$$^{66}$Zn values that also push the average to a lighter average value of –0.15$\pm$0.44 ‰. We can relate these variations to chondrule/CAI contaminations or heterogeneities of the matrix. The positive endmember $\updelta$$^{66}$Zn value of the ICM (0.14$\pm$0.12 ‰) coincides with a large sulfide grain in the Bmx2 sample, whereas the more negative $\updelta$$^{66}$Zn values correlate with observed chondrule contaminations. Collectively, pure ICM has a light Zn isotope composition relative to the FGRs and to the bulk CV chondrite and the heaviest FGR (Ch2mx, Table \ref{tab:CrZn}) has a $\updelta$$^{66}$Zn value that is within error of CV and CI bulk chondrites. In addition to the matrix, we have sampled a large fraction of a chondrule core and two whole chondrules (i.e., combined core and igneous rim) to verify the Zn isotope compositions reported for singularly small samples from Leoville chondrules \citep{van_kooten_zinc_2019}. These chondrule cores are very Zn-depleted and previous data was obtained on only 1-5 ng of Zn, thus, involving large blank contributions of about 20 \%. The Zn isotope analyses of a larger Leoville chondrule core (10 ng Zn) in this study shows that, after blank correction, the large core composition ($\updelta$$^{66}$Zn = –0.94$\pm$0.12 ‰) is indistinguishable from the previous measurements ($\updelta$$^{66}$Zn = –0.81$\pm$0.18 ‰, \citealp{van_kooten_zinc_2019}). Moreover, the bulk chondrule compositions measured here are in agreement with previous analyses of Allende bulk chondrules \citep{pringle_origin_2017}. Furthermore, we analyzed an Al-rich chondrule, which in contrast to the other chondrule cores has a positive $\updelta$$^{66}$Zn value of 0.10$\pm$0.12 ‰.

\begin{figure}[bt!]
\centering
\includegraphics[width=0.9\textwidth]{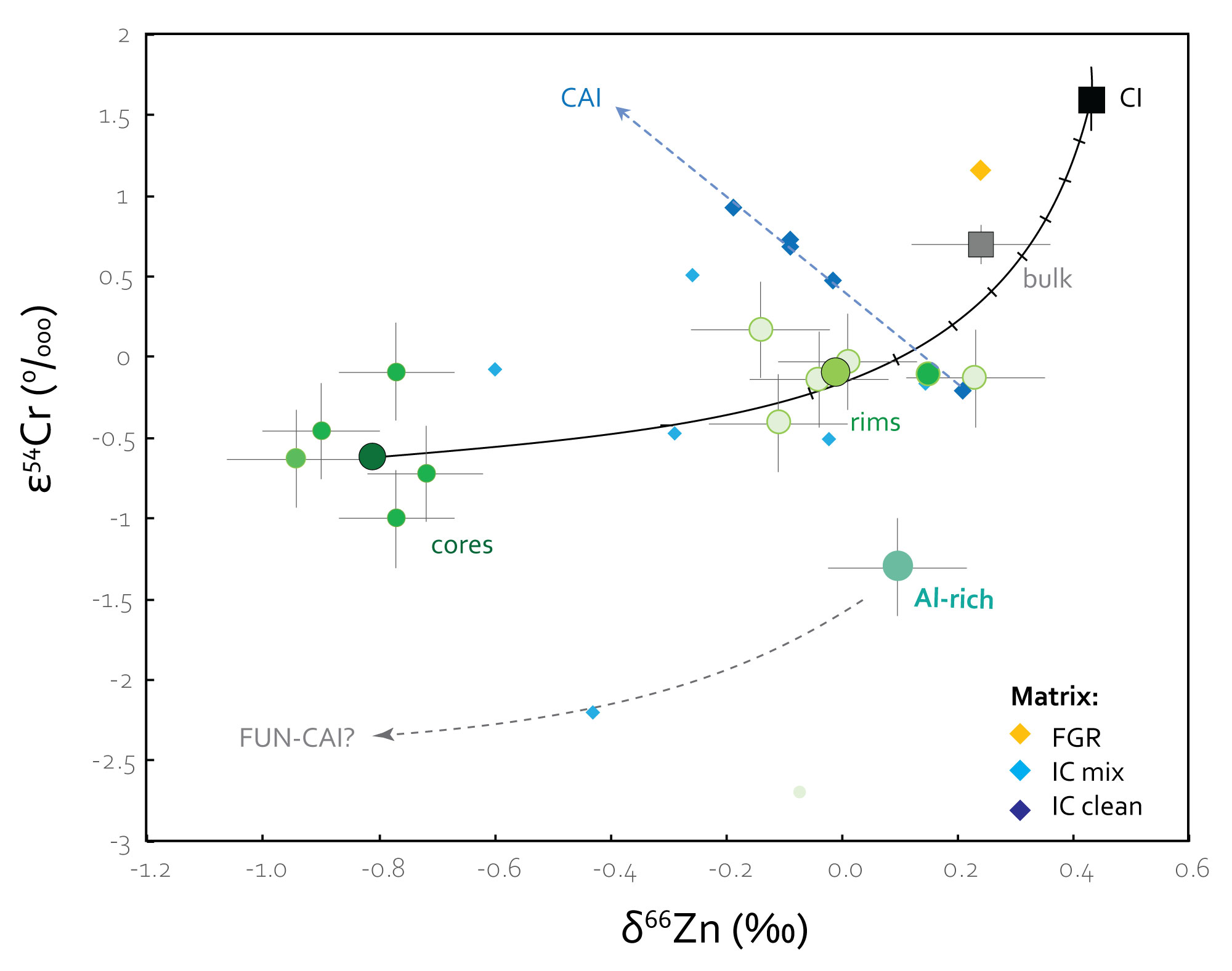}
\caption{Plot of mass-independent $\upepsilon$$^{54}$Cr versus mass-dependent $\updelta$$^{66}$Zn values of Leoville components, reflecting their potential heritage and precursor materials. The averaged isotope compositions of chondrule cores and rims (bigger spheres) are also shown. The spheres represent chondrules, the squares reflect bulk chondrites (grey = CV) and the diamonds represent the matrix. A mixing line is drawn with CI chondrites (Cr = 2590 ppm and Zn = 310 ppm, \citealp{lodders_solar_2003}) and Leoville chondrule cores (Cr = 3000 ppm and Zn = 50 ppm) as endmembers. We note that the curvature of the mixing line is mostly dependent on the estimated Zn content of the chondrule cores (including metal rim), which is currently unknown. The tick marks represent fractions of 0.1. Notably, most Leoville components fall on this mixing line, including the chondrule rims, the ICM, the CV bulk and the FGR. The clean ICM is a mixture between CI-like material, chondrules and CAIs. This suggests that variations in $\updelta$$^{66}$Zn are not only related to volatile loss processes, but are also related to their precursor materials. The Al-rich chondrule Bch6 and one ICM sample do not fall on this mixing line, suggesting that some of this material has FUN-CAIs as precursor, rather than canonical CAIs.). 
\label{fig:CrZn}}
\end{figure}

\section{Discussion} \label{sec:discussion}
\subsection{The effect of secondary alteration} \label{sec:alteration}
Prior to making an attempt to understand the origin of chondritic components through their (nucleosynthetic) isotope systematics, it is of critical importance to recognize the effects of elemental and isotopic exchange between these components during secondary alteration (e.g., thermal metamorphism and/or aqueous alteration). Although the use of nucleosynthetic isotopes is often perceived as providing a means to assess the heritage of precursor materials to thermally processed chondritic components independent of secondary processes, this method is not assumption free. Notably, isotopic fractionation occurs during elemental exchange between chondritic components and the magnitude and kinetics of this fractionation are subject to the redox conditions, fluid composition and temperature of alteration. Given that alteration typically is a low temperature process where equilibrium processes dominate, the resulting  mass-dependent isotope fractionation is not necessarily appropriately accounted for by the typical kinetic mass fractionation laws applied to correct stable isotope data obtained by mass spectrometry. Indeed, mass-dependent isotope fractionation at low-temperature conditions such as experienced during moderate secondary alteration on carbonaceous chondrite parent bodies theoretically and empirically result in large isotopic variations. Especially for the redox-sensitive chromium, extreme fractionations up to 7 ‰ (at 298 K, per 1 amu) have been predicted \citep{schauble_theoretical_2004}. Since for every 0.1 ‰ of fractionation can generate a potential bias on the  mass-independent $\upepsilon$$^{54}$Cr of 0.04, the total effect of inappropriately corrected equilibrium isotope fractionation between chondritic components can potentially result in epsilon-level shifts in the $\upepsilon$$^{54}$Cr value \citep{schiller_precise_2014,van_kooten_magnesium_2017}. Given that Cr is highly mobile even at the most early stages of alteration, which is highlighted by the fact that the chromium content and distribution in olivine phenocrysts from ordinary and CV chondrite chondrules has been used to determine the degree of thermal metamorphism \citep{grossman_onset_2005}, reinforces that extreme caution is needed when utilizing the Cr isotope composition of individual chondrules and matrix to make inferences about their genetic relationships.\\

What is unambiguous from our Cr isotope measurements of the Leoville chondrule cores is that their $\upepsilon$$^{54}$Cr values are significantly lower than those reported for chondrules from more altered CV chondrites. With increasing degree of thermal metamorphism (Leoville $<$ Vigarano $<$ NWA 3118 $\sim$ Allende) the average chondrule $\upepsilon$$^{54}$Cr value approaches that of the bulk CV chondrite (Fig. \ref{fig:e54Cr}). This suggests that the Cr isotope signatures of chondrules and matrix are progressively homogenized during thermal metamorphism. The major implications of these results are that 1) the previously reported range of $\upepsilon$$^{54}$Cr values is to some extent the result of secondary alteration \citep{kadlag_cr_2019,olsen_magnesium_2016}, 2) the $\upepsilon$$^{53}$Cr values from chondrules used for isochrons in the Mn-Cr decay system need to be reassessed according to their level of alteration \citep{zhu_chromium_2019} and 3) other isotope systems such as Fe, W and Mo that are redox-sensitive may also need to be revisited. For example, stable Fe isotope variations between chondrules and matrix from the moderately altered CM chondrite Murchison have been interpreted in favor of chondrule-matrix complementarity \citep{hezel_fe_2018}. Furthermore, complementary mass-independent isotope variations of W and Mo have been proposed between chondrule and matrix from the altered CV chondrite Allende \citep{budde_molybdenum_2016}. We suggest that the Leoville chondrule cores, igneous rims and matrix provide accurate constraints on the origin and formation mechanisms of these components through their Cr and Zn isotope signatures, respectively. We note, however, that even Leoville is not completely unaltered as some of its metal is sulfurized \citep{van_kooten_unifying_2019}. A detailed investigation of the CV chondrite metal (isotope) composition may improve our understanding of the contribution of metal to the total mass balance of the carbonaceous chondrites and the effect of oxidation on the composition of the matrix \citep{van_kooten_unifying_2019}. For example, leaching of the metal by oxidation can bring siderophile elements such as W and Mo into the matrix and sulfurization can concentrate Zn in secondary products. 

\subsection{The origin of CV chondrite components} \label{sec:origin}
The first order observation emerging from our work is that the chondrule-rim systems in the pristine Leoville CV chondrite span a range of $^{54}$Cr compositions that is similar to that observed for the Solar System’s planets and asteroidal bodies, namely from ureilite-like to CI-like signatures. The onion-ring like structure of individual chondrules record an increasingly $^{54}$Cr-rich composition from core to outer rim. This observation firmly establishes that NC and CC material coexisted in space and time, which has far reaching implications for understanding mass transport processes and the origin of chondritic components that we discuss below.\\

In detail, the $\upepsilon$$^{54}$Cr values of Leoville chondrule cores define a relatively narrow and negative range compared to previous measurements \citep{kadlag_cr_2019,olsen_magnesium_2016}. Interestingly, the range of $\upepsilon$$^{54}$Cr values of the cores corresponds to that of inner Solar System materials, such as eucrites, angrites, ordinary chondrites \citep{trinquier_widespread_2007} and ureilites ($\upepsilon$$^{54}$Cr = –0.91$\pm$0.15 ‱;  \citealp{zhu_chromium_2020}) of which the latter are suggested to represent an endmember composition of planetary bodies that accreted in the terrestrial planet region \citep{schiller_isotopic_2018}. The ureilite parent body likely formed $<$1 Myr after CAI formation \citep{van_kooten_magnesium_2017} and its low bulk $\upepsilon$$^{54}$Cr value is interpreted to reflect: 1) a formation location in the asteroid belt before injection of $^{54}$Cr-rich material into the protoplanetary disk \citep{goodrich_origin_2015} or 2) formation close the Sun, where its $\upepsilon$$^{54}$Cr value is the result of removal of the $^{54}$Cr-rich carrier, perhaps via thermal processing, from the dust \citep{larsen_evidence_2011}. However, since astrophysical models of injection of supernova-derived dust to the disk show a very low probability of explaining the high abundance of supernova-derived, short-lived radionuclides (and by extension $^{54}$Cr) \citep{ouellette_injection_2010}, it is more likely that the low $\upepsilon$$^{54}$Cr values of the chondrules reflect a formation location close to the protosun. Either way, the Cr isotope composition of Leoville chondrule cores establishes that these objects initially originated in the terrestrial planet formation region as opposed to the outer Solar System, which is characterized by positive $\upepsilon$$^{54}$Cr values \citep{van_kooten_isotopic_2016}. The oxygen isotope systematics of the least altered CV chondrite chondrules support relatively dry conditions during their final melting event \citep{hertwig_formation_2018}, in agreement with an inner Solar System origin of these chondrules. Furthermore, the Zn isotope compositions of the chondrule cores are remarkably similar (Table \ref{tab:CrZn}, \citealp{van_kooten_zinc_2019}). Zn stable isotope fractionation during chondrule formation is thought to occur either through volatile loss from the chondrule or partitioning of Zn to the metal/sulfide melt and subsequent metal-silicate separation. Hence, our data suggest a very similar formation mechanism for these chondrules from the same spatiotemporal reservoir. Exceptions from this are the Al-rich chondrules Bch6 and Bch3. Al-rich chondrules are thought to have formed from CAI-like precursor materials \citep{zhang_origins_2014}, which have very negative $\updelta$$^{66}$Zn values \citep{luck_zn_2005}. In agreement with the presence of a CAI precursor, Bch3 has a very high $\upepsilon$$^{54}$Cr value (Zn isotope composition not measured). Bch6 is distinct from this in that it has both  the lowest $\upepsilon$$^{54}$Cr value measured here and a positive $\updelta$$^{66}$Zn value, unlike the more typical type I chondrules (Fig. \ref{fig:CrZn}) and contrary to the prediction based on a CAI-precursor. However, it is also possible that Bch6 inherited a FUN-CAI precursor, which exhibit very negative $\upepsilon$$^{54}$Cr values \citep{holst_182hf182w_2013}.\\

The igneous rims have higher $\updelta$$^{66}$Zn and $\upepsilon$$^{54}$Cr values compared to the chondrule cores. These rims are thought to represent chondritic dust that is thermally processed during transient heating events in the disk prior to accretion onto chondrule cores as volatile depleted forsterite-bearing grains \citep{van_kooten_zinc_2019}. Subsequently, the grains react with surrounding volatile-rich gas, which progressively enriches the igneous rims in heavy Zn isotopes as well as moderately volatile elements. This results in a near-chondritic elemental composition of the igneous rims, as well as forsterite relict grains overgrown by low-Ca pyroxene co-existing with troilite/metal assemblages. Based solely on the Zn isotope data igneous rims may have formed concurrently to chondrule cores. However, the $^{54}$Cr-enrichment of the rims suggests that they reflect a separate and less energetic heating events than those that formed chondrule cores. During this second heating event, increasingly $^{54}$Cr-rich material was added to the chondrule growing reservoir. The higher $\upepsilon$$^{54}$Cr signature of chondrule rims relative to their cores cannot result from inappropriately corrected mass-dependent equilibrium isotope fractionation, since estimated temperatures of igneous rim formation by sintering (T $<$ 1000-1200 K; \citealp{jones_thermal_2018,rubin_chondrules_1987}) are too high to cause significant fractionation during gas-melt interaction between CrO (liquid) in chondrule silicates and CrO$_{2}$ (gas) and the expected fractionation is predicted to result in a lighter final silicate \citep{sossi_volatile_2018}. Hence, the $^{54}$Cr enrichment of the rims over the cores must be the product of addition of $^{54}$Cr-rich material to the chondrule forming reservoir. This also implies that the stable isotope enrichment in $^{66}$Zn is not necessarily the product of gas-melt interaction, but may also reflect a change in the chemical nature of the precursor materials. The presence of the igneous rims reflects reheating of chondrules at relatively low temperatures in a less energetic disk environment with lower gas densities, which could have been in 1) an older disk and/or 2) at larger orbital distances.
The fine-grained dust rims that accreted around the chondrules have a very homogeneous CI-like chemical composition \citep{van_kooten_unifying_2019} and corresponding high $\upepsilon$$^{54}$Cr and $\updelta$$^{66}$Zn values. We note that even though we only measured the Cr isotope composition of a single FGR, the mass balance of all CV components requires FGRs to have CI-like Cr isotope signatures since chondrules and intra-chondrule matrix have similar or lower $\upepsilon$$^{54}$Cr values than bulk CV chondrites and addition of refractory solids is insufficient to explain the bulk value (Appendix C). Collectively, the nucleosynthetic Cr isotope systematics of the FGRs agree with being derived from pristine CI-like dust and mass-dependent Zn isotope systematics suggest that the FGRs did not experience significant volatile gain or loss. This is in agreement with the model that matrix and chondrules are not complementary to each other \citep{van_kooten_unifying_2019}. Rather, our Cr isotope data suggest that the CV chondrite reservoir was increasingly enriched in CI-like dust from the outer Solar System. \\

Finally, the Cr isotope composition of the pure intra-chondrule matrix is indistinguishable from that of bulk CV chondrites (Fig. \ref{fig:e54Cr}), whereas the Zn isotope composition is typically lighter. This appears to be the result of addition of CAI-like material to the ICM (Fig. \ref{fig:CrZn}). Overall, petrological and isotopic characteristics suggest that the ICM represents a mixture of chondrule, CAI and matrix fragments. The visually ‘contaminated’ ICM including chondrule fragments typically has more negative Cr and Zn isotope values, with one of the matrix samples being significantly more negative than the chondrules. Similar to the Al-rich chondrule Bch6, this may be the result of addition of FUN-CAI-like material. Collectively, the nature of the ICM implies that this material is not by itself a primary nebular product but reflects a mixture of all CV chondrite components. This, in turn, indicates that the CV ICM is a parent body feature that formed during accretion and physical erosion of the CV chondrite components. Hence, although the fine-grained dust within the ICM may still retain some primary features \citep{haenecour_presolar_2018}, the overall composition is not primary and this material should not be used as such. 

\subsection{Massive outward transport of inner Solar System chondrules} \label{sec:transport}
The current paradigm proposes that carbonaceous chondrites, including CV chondrites, formed in the outer Solar System, beyond the accretion region of Jupiter. Convincing arguments in favor of this model include the petrologic and isotopic dichotomy of bulk NCs and CCs (e.g., Mo, O, Cr, Ca, Ti; \citealp{budde_molybdenum_2016, schiller_isotopic_2018,warren_stable-isotopic_2011}). Additionally, previous Cr isotope measurements of individual CV chondrules have been interpreted to indicate that CV chondrites accreted in a spatially distinct reservoir from inner Solar System materials \citep{olsen_magnesium_2016}. At face value, our results suggest that CV chondrites do not contain outer Solar System derived chondrules but represent hybrid bodies consisting of inner Solar System chondrule cores and outer Solar System derived CI-like dust. If CV chondrites are representative of other chondrule-bearing carbonaceous chondrites, this implies that chondrules did not form beyond the accretion region of Jupiter. The exception from this are the metal-rich carbonaceous chondrites (i.e., CR, CB and CH), for which the chondrules contain exclusively $^{54}$Cr-rich outer Solar System signatures. These chondrites likely formed beyond the accretion region of the gas giants \citep{van_kooten_role_2020}. Hence, the major chondrule-forming factory may have been restricted to the inner Solar System such that outer Solar System chondrules only represent a minority of objects, perhaps formed by impacts (i.e., CH and CB, \citealp{krot_young_2005}), bow shocks (i.e., CR, \citep{morris_chondrule_2012,van_kooten_isotopic_2016}), or in the vicinity of the giant planets \citep{bodenan_can_2020}.\\

\begin{figure}[ht!]
\centering
\includegraphics[width=1\textwidth]{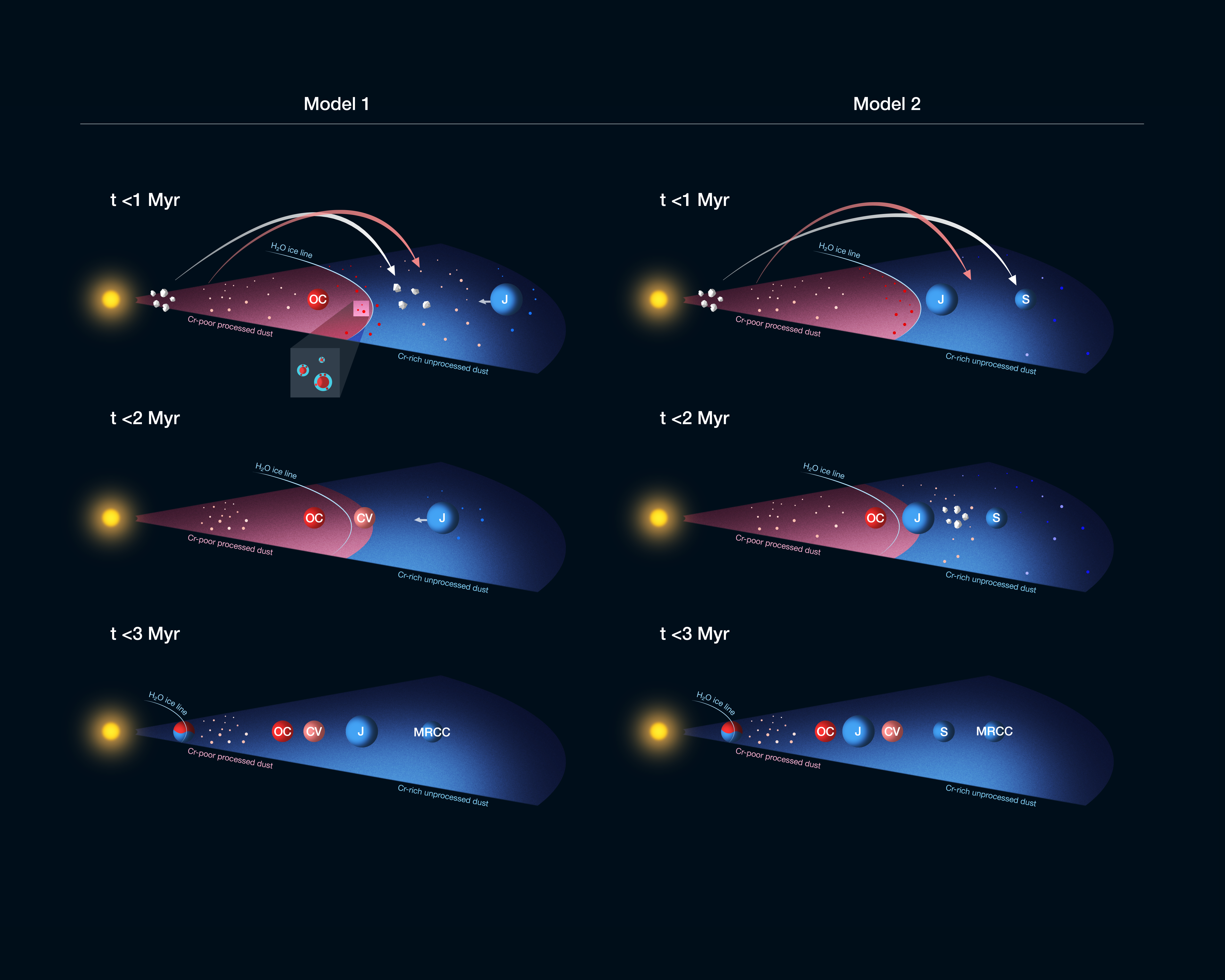}
\caption{Schematic representation of a protoplanetary disk showing two chondrite accretion scenarios, with Jupiter as a barrier separating NC and CC (model 2) and without Jupiter (model 1). At t$<$1 Myr, CV (pink spheres) and OC chondrule cores (red spheres) form in the inner Solar System in different oxidizing environments. OC chondrules form close to the water ice sublimation front, where they rapidly accrete FGRs \citep{krot_microchondrules_1997} within the ice line in a high dust-to-gas region (Ida and Guillot, 2016). CV chondrule cores and CAIs are transported outwards to (model 2) beyond Jupiter's accretion region at 3-5 AU or (model 1) outside the ice line, but within Jupiter's initial core accretion region at $>$30 AU \citep{oberg_jupitertextquotesingles_2019}. CR chondrules (blue spheres) form beyond the gas giant orbits \citep{van_kooten_role_2020}. Earth's initial feeding zone consists of CV-like type I chondrules surrounded by metal rims and with ureilitic Cr isotope signatures. This is the metal that forms Earth's core and leaves the mantle depleted in Fe \citep{schiller_iron_2020}. Model 1: At t$<$2 Myr, CV CAIs and chondrules accrete CI-like FGRs outside of the stagnant water ice line (see text for further details) and grow to form planetesimals \citep{ros_ice_2013}. CAIs and other mm-sized to cm-sized objects are caught in the rapid pile-up of material in planetesimals and a significant fraction of these objects do not pass the ice line barrier. In both models: At t$<$3 Myr, as the accretion of NC and CCs is finalized, the ice line migrates further inwards and the chondrules in Earth's feeding zone can accrete CI-like dust rims. These dust rims accrete to Earth's final mass during pebble accretion and result in the Earth having a CI-like Fe isotopic composition of its mantle \citep{schiller_iron_2020}. The metal-rich carbonaceous chondrites accrete beyond Saturn's (model 2) or Jupiter’s (model 1) orbit.
\label{fig:cartoon}}
\end{figure}

Previous studies have proposed that CC chondrules are the product of outer Solar System precursor dust mixed to various degrees with $^{54}$Cr-rich refractory inner Solar System grains such as spinels and/or another unknown carrier \citep{gerber_mixing_2017,williams_chondrules_2020}. In this model, CC chondrules are generally viewed as outer Solar System products with an incorporated ‘nugget effect’. In detail, CC chondrule formation is suggested to be in situ (i.e., where the CC chondrites accreted) and ascribed to mixing, agglomeration and subsequent heating of inner and outer Solar System dust. If correct, this model predicts that a range of $^{54}$Cr values should be observed for CV chondrules. Although a compositional range has been reported for altered CV chondrites such as Allende and NWA 3118, we attribute this observation to secondary alteration (Fig. \ref{fig:e54Cr}). Our data based on the pristine Leoville CV chondrite firmly establishes that chondrule cores are $^{54}$Cr-depleted and, as such, cannot have formed from outer Solar System precursors. An important implication of our results is that massive outward transport of inner Solar System chondrules to the accretion regions of carbonaceous chondrites must have taken place during the early evolution of the protoplanetary disk (Fig. \ref{fig:cartoon}, model 2). Thus, outward mass transport was not limited to refractory inclusions and $\upmu$m-sized dust but also included the most abundant chondrite component, namely chondrules.\\

The U-corrected Pb-Pb age dating of inner Solar System chondrules support the hypothesis that these objects formed early and contemporaneously \citep{bollard_early_2017,connelly_absolute_2012}. Indeed, absolute chronology of individual ordinary chondrite chondrules indicate that the bulk of these objects formed within 1 Myr of Solar System formation. As discussed above, Leoville chondrule cores were likely formed in the inner Solar System and presumably in the accretion region of ureilites, relatively close to the protosun. Cr, O and Zn isotope data of the most unaltered CV chondrule cores suggest that these components formed from a single population (\citealp{hertwig_formation_2018}, this study). This is in agreement with the model of thermal processing, where the precursor material of these cores consisted of $^{54}$Cr-depleted thermally processed dust complementary to the $^{54}$Cr-rich gas from which the CV CAIs originally condensed \citep{larsen_evidence_2011}. Hence, similar to ordinary chondrite chondrules, the formation of CV chondrules may have overlapped with that of CAIs and was restricted to the first Myr of Solar System formation \citep{bollard_early_2017,connelly_absolute_2012}. We note that individual Pb-Pb ages of CV chondrules are rare and only exist for the moderately altered Allende \citep{connelly_absolute_2012}, which agreeably have ages indistinguishable from CAI formation. These age-dated Allende chondrules were selected based on their low degree of alteration and coincidentally also have negative $^{54}$Cr isotope signatures relative to Earth, in agreement with our results and interpretations.\\

Outward mass transport of solids is modeled to be most efficient in the earliest evolutionary stage of the protoplanetary disk, either via stellar and disk outflows or, alternatively, during viscous expansion of the disk \citep{shu_toward_1996,cuzzi_blowing_2003,ciesla_outward_2007,ciesla_distributions_2010,haugbolle_probing_2019}. Thus, outward mass transport of inner Solar System chondrules to the outer Solar System likely occurred prior to the formation of Jupiter, that is, within the first Myr of disk evolution. Once CV chondrules were transported and stored at the pressure trap created by a planetary gap opened by Jupiter, the chondrule cores were coated by a progressive addition of CI-like dust (Fig. \ref{fig:cartoon}, model 2). The chondrules probably experienced brief heating events that formed the igneous rims in the outer Solar System. Finally, after the last coating of fine-grained dust, the CV chondrules grinded together within rubble pile asteroids located at Jupiter’s pressure trap \citep{eriksson_pebble_2020}, thereby forming a second generation of dust: the intra-chondrule matrix. These products cemented together forming the CV chondrites. Importantly, Jupiter’s pressure trap does not prevent fine-grained dust to be transported through the gap \citep{haugbolle_probing_2019,weber_characterizing_2018}. This implies that if coating of chondrules by CI-like dust occurred beyond Jupiter’s orbit, the same process was experienced by chondrules present in the terrestrial planet region. Hence, if Earth indeed accreted its main mass through pebble accretion, and specifically by chondrules \citep{johansen_growth_2015}, Earth would have been progressively enriched by CI-like material (Fig. \ref{fig:cartoon}, model 2). This is in agreement with the Fe isotopic data that predict a significant part of Earth’s mass should be CI-like \citep{schiller_iron_2020}. Although the formation of Jupiter may have aided the storage of significant amounts of mm-sized solids such as inner Solar System chondrules and refractory inclusion beyond its orbit, Jupiter did not prevent the mass transfer of outer Solar System CI-like dust to the inner Solar System. Thus, in contrast to current belief, the early formation of Jupiter had limited impact on the compositional gradient that exist between inner and outer Solar System bodies.

\subsection{A potential inner Solar System accretion region for CV chondrites?} \label{sec:icelines}
The previous section assumed Jupiter to pose as a barrier that prevented the macroscopic components from the CC (i.e., chondrules and CAIs) from drifiting into the inner Solar System and, based on this assumption, that all carbonaceous chondrites accreted beyond the orbit of Jupiter. However, considering our new results and the latest developments in astrophysical models, we explore whether the accretion region of CV chondrites and, by extension, that of other CCs is required to be in the outer Solar System. Focusing solely on the Cr isotope compositions of unaltered CV chondrite components, then it is not necessary to accrete CV chondrites in the outer Solar System. Indeed, the observation that the $^{54}$Cr-poor compositions of the CV chondrule cores are mantled by increasingly more $^{54}$Cr-rich rims is consistent with the progressive addition of CI-like material to an inner Solar System accretion region. If this is correct, then the other physicochemical constraints on the formation regions of NCs versus CCs should be met as well. Specifically, the observed isotopic and petrological dichotomy between these chondrite groups needs to be accounted for. For this purpose, we compare constraints from ordinary chondrites (NC) to those of CV chondrites (CC).\\

The make-up of ordinary chondrites may not be very different from CV chondrites. The sizes \citep{friedrich_chondrule_2015} and Cr isotope compositions of OC chondrules overlap with those of CV chondrule cores (Fig. \ref{fig:e54Cr}). Oxygen isotope analyses of the least altered OC and CV chondrules suggest that these chondrules formed in different disk reservoirs and/or at different times. The $\Delta^{17}$O values of Semarkona, Bishunpur and Krymka (LL3.00-3.1; \citealp{kita_high_2010}) chondrules define a narrow range of $\sim$0.7 ‰, whereas Kaba chondrules (CV3.1$_{ox}$) have $\Delta^{17}$O values below the terrestrial fractionation line and on the PCM slope between –8 and –4 ‰ \citep{hertwig_formation_2018}, although a fraction of OC chondrules has $\Delta^{17}$O values similar to CV chondrules \citep{hertwig_formation_2018}. A plausible interpretation of the $\Delta^{17}$O systematics of OC chondrules is that these objects originated in an oxidizing environment, where $^{16}$O-depleted gas interacted with silicate dust of similar isotope composition further away from the protosun. Alternatively, the relatively high $\Delta^{17}$O values of OC chondrules reflect a higher silicate/dust ratio in the chondrule forming region \citep{tenner_oxygen_2015}. Such environments are predicted to exist at the sublimation front of the water ice line (Fig. \ref{fig:cartoon}, model 1), which would have been located near the current asteroid belt at the time of ordinary chondrite accretion \citep{dodson-robinson_ice_2009}. Given that our $^{54}$Cr isotope data indicate an inner Solar System origin for CV chondrules, a possibility is that the lower $\Delta^{17}$O values of these objects reflects a formation region sunwards of the OC chondrule forming region (Fig. \ref{fig:cartoon}, model 1). Within the FGRs of OCs, microchondrules are observed that match the petrology and oxygen isotope systematics of their host chondrules \citep{dobrica_microchondrules_2016,krot_microchondrules_1997}. Therefore, the accretion of fine-grained dust around these chondrules must have been practically concurrent with chondrule formation in OCs and, hence, also occurred near the water ice line (Fig. \ref{fig:cartoon}, model 1). These early formed mm-sized chondrules and their rims would have drifted sunward until they encountered either a physical trap (i.e., a pressure bump) or a region of increased pebble density (i.e., a pile up zone) to store/accrete them, since the relatively rapid rate of inward radial drift would not allow them to remain in the disk long enough to form planetesimals otherwise \citep{desch_formulas_2017}. A pile-up of pebbles has been theorized to exist at the sublimation front of the water ice line by accumulation of deaccelerating sublimated silicates, which are piled-up to form regions of high dust-to-gas ratios and, finally, gravitationally collapse to form planetesimals \citep{ida_formation_2016}. Collectively, this would suggest early formation of OC chondrules and rapid accretion of their parent bodies. This is in agreement with the modelled accretion ages of ordinary chondrites $<$1.8 Myr after CAI formation (earlier if assuming a heterogeneous distribution of $^{26}$Al, Appendix D) \citep{doyle_early_2015}. \\

The final accretion region of CV chondrites and, by extension, other CCs is inferred to be outwards from the ordinary chondrite accretion region, since the CV parent bodies accreted later \citep{doyle_early_2015} and contain a larger fraction of pristine CI-like matrix. Materials that are initially transported outwards invariably drift inwards towards the central star unless they can overcome the radial drift barrier, usually by decoupling from the gas by changing the solid-to-gas ratio in the disk \citep{gonzalez_self-induced_2017}. Although various so-called ‘pressure traps’ have been proposed in the past (i.e., long-lived vortices etc.), the most popular barrier to separate NCs from CCs is invoked by the opening of a gap associated with giant planet formation \citep{kruijer_age_2017}. In detail, the lack of CAIs in the terrestrial accretion region and the distinct isotopic make-up of concurrently accreting NCs versus CCs is attributed to the formation of Jupiter and the resulting separation of reservoirs \citep{haugbolle_probing_2019}. The isotopic dichotomy is also observed for the iron meteorites that are linked to NCs and CCs by their Mo isotope signatures \citep{kruijer_age_2017}. The Hf-W model ages of these irons and the chondrites suggest that a barrier should have been in place at $<$1 Myr after CAI formation. However, a potential problem with using Jupiter as a barrier is that its chemical composition does not allow its core to have formed at the proposed location of 3–5 AU. In fact, an initial accretion location at $>$30 AU is inferred from the position of N$_{2}$ and noble gas ice lines as well as the asymmetry in its two populations of Trojan asteroids  \citep{oberg_jupitertextquotesingles_2019,pirani_consequences_2019}. This early and distal accretion is followed by $\sim$2 Myr of migration until final accretion of its envelope at 5 AU. If Jupiter formed at $>$30 AU, this requires that CAIs and chondrules were transported outwards to such distances, which has been proposed to occur through jets and outflows \citep{haugbolle_probing_2019} or by outward turbulent diffusion \citep{ciesla_outward_2007}. In the former scenario, CAIs that are not accreted by the protosun are ejected as part of the outflow and either lost to the interstellar medium or distributed in the protostellar envelope as part of the entrainment and mixing of envelope and jet material \citep{haugbolle_probing_2019}. They then return to the protoplanetary disk during the general accretion of the envelope in the protostellar phases on a timescale of 500 kyr \citep{haugbolle_probing_2019}. In outward diffusion models developed by \citet{ciesla_outward_2007} and later models including particle growth during transport \citep{misener_tracking_2019}, materials can be exchanged over a radial span of 5–20 AU on timescales of only $\sim$10$^{5}$ yr. \citet{misener_tracking_2019} suggest that a barrier must be in place within these spatiotemporal parameters to prevent mixing between NC and CC reservoirs. Therefore, if CAIs are transported radially by viscous spreading, either Jupiter did not originate at $>$30 AU or Jupiter is not the cause for the NC – CC dichotomy.\\

An alternative possibility to Jupiter is that the water ice line acted as the barrier that separates the NC and CC reservoirs. Ice lines have been proposed to initiate rapid pile-ups of materials that form planetesimals and are related to the sublimation \citep{ida_formation_2016} and condensation \citep{drazkowska_planetesimal_2017,hyodo_formation_2019,schoonenberg_planetesimal_2017,schoonenberg_lagrangian_2018} of ice. In detail, while on the inside of the ice line an increased solid-to-dust ratio is related to the deacceleration of sublimated particles, the outside of the ice line is defined by outward diffusion and condensation of water and resulting streaming instabilities. As we outline below, certain feasible disk models allow for the ice line to remain relatively stagnant and positioned within the asteroid belt throughout the accretion timescales of chondrite parent bodies (assuming internal heating of the disk, \citealp{bethune_electric_2020}). During the lifetime of the protoplanetary disk, the accretion rate to the central star is expected to decline \citep{hartmann_accretion_1998} and, as the gas density decreases, the water ice line migrates inwards. As a result, within a $<$3 Myr period the ice line is expected to be around 1 AU in most disk models \citep{dodson-robinson_ice_2009,oka_evolution_2011}. CV chondrites have modelled accretion ages of $<$2.6 Myr after CAI formation \citep{doyle_early_2015} (earlier depending on the distribution of $^{26}$Al in the disk, Appendix D) and, thus, the separation of macroscopic NC and CC components is expected to last at least that long. Hence, if the water ice line indeed acted as barrier, the rate of inward migration must have been slow enough to accommodate this separation. The speed of ice line migration is highly dependent on the accretion rate onto the central star. This rate is highest during the class I stage (10$^{5}$ yrs) where the protostar is embedded in the molecular cloud and the protoplanetary disk is replenished through infall from the envelope \citep{hartmann_fu_1996}. During this stage, FU Orionis type outbursts can shift the ice line far outwards, with retention rates of 10-100s yrs. The accretion rate declines over time/with disk age, but this decline is not proven to be constant and the uncertainty on its progression is more than an order of magnitude \citep{hartmann_accretion_1998,li_lifetimes_2016}. Indeed, for some stars the accretion rate may still be 10$^{-8}$ $M_\odot$/y after 3 Myr, which would suggest that the water ice line could be relatively stagnant during that time. Recent models also show that the position of the ice line is dependent on other disk parameters, such as the radial distribution of the turbulent viscosity parameter $\upalpha$ in the disk and the growth rate of chondrules into asteroids \citep{kalyaan_effect_2019}. These parameters influence the balance between the outward diffusion of water vapor and the growth of icy pebbles outside of the ice line. These models show that it is possible to form the distribution of water-poor and water-rich asteroids in situ in the asteroid belt \citep{alexander_provenances_2012}, without a need for the scattering of asteroids by a Grand Tack \citep{walsh_low_2011} or for the fossilization of the water ice line through the formation of Jupiter \citep{morbidelli_fossilized_2016}. This ice line fossilization was invoked to prevent icy particles from populating the terrestrial planet region \citep{morbidelli_fossilized_2016}. If Jupiter indeed formed at large orbital distances, we consider the possibility that chondrite parent bodies formed in situ at their current positions in the asteroid belt and were chemically separated via the water ice line that remained stagnant for a period $<$2.6 Myr. The efficiency of this chemical separation depends on the relative accretion timescales of NC and CC as well as the rate of planetesimal accretion at the ice line. It is likely that at least some material will pass through the ice line. For example, chondrules have been found in enstatite chondrites with an outer Solar System Cr isotope signature \citep{Zhu_2020} and rare CAIs in ordinary chondrites have similar oxygen isotope systematics as those found in CC \citep{Huss2001}. 

\section{Conclusions} \label{sec:conclusions}
Our detailed Cr and Zn isotope investigation of the relatively unaltered carbonaceous chondrite Leoville and its components sheds light on the origin and mass transport of chondrules and matrix in the protoplanetary disk. The first order observation emerging from our work is that Leoville’s chondrule-rim systems span a range of $^{54}$Cr compositions that is similar to that observed for the Solar System’s planets and asteroidal bodies, namely from ureilite-like to CI-like signatures. The onion-ring structure of individual chondrules record an increasingly $^{54}$Cr- and $^{66}$Zn-rich composition from the chondrule core to the outer rim. These data show that the observed Cr isotopic range in chondrules from more altered CV chondrites is the result of chemical equilibration between chondrules and matrix during secondary alteration. We propose two scenarios that could account for the hybrid isotopic nature of CV chondrites, subject to the radial distance at which Jupiter originated. First, if proto-Jupiter opened a disk gap within $<$1 Myr after CAI formation at 3-5 AU, CV chondrite accretion is best explained by massive outward transport of chondrules combined with an inward flux of CI-like dust. These components were then trapped beyond Jupiter and cemented together. In this interpretation, the CI-like fine grained dust that accreted to the CV chondrules was also transferred inwards across the gap and into the terrestrial planet region. Bidirectional transport of chondrules in outflows outwards and fine grained matrix strongly coupled to the accreting gas inwards take in to account the full dynamics in the protosolar environment, and nuances the view of Jupiter as an efficient barrier for solids in the NC and CC reservoirs. Second, since our data do not require an outer Solar System accretion region for CV chondrites, we consider an alternative model in which the water ice line, rather than Jupiter, acted as a barrier between NC and CC reservoirs. This consideration is rationalized by recent models that propose Jupiter originated from beyond 30 AU. Indeed, a range of astrophysical conditions within the protoplanetary disk allow for a relatively stagnant ice line within the timeframe of chondrite accretion. However, this scenario requires further verification via numerical simulations that explore the efficiency of ice line pile-ups as well as the transport and retention time of refractory solids within the protoplanetary disk. 
Finally, we note that astrophysical models show that water-rich planetesimals at the ice line may accrete rapidly, independent of the role of Jupiter in modulating mass transport between the inner and outer Solar System. This implies that two populations of CCs should exist. The first one formed beyond Jupiter and contains chondrules and CAIs. The second one, if all chondrules and CAIs were blocked by Jupiter, contains only fine-grained dust coated by ice. Hence, CI chondrites, with their unique petrology relative to other CCs, could have formed at the ice line within Jupiter’s orbit. 

\acknowledgments
We thank Alexander Krot for valuable discussion on oxygen isotope systematics in chondrules. This project has received funding from the European Union’s Horizon 2020 research and innovation programme under the Marie Sk\l{}odowska-Curie Grant Agreement No 786081 to E.K. F.M. acknowledges funding from the European Research Council under the H2020 framework program/ERC grant agreement no. 637503 (Pristine) and financial support of the UnivEarthS Labex program at Sorbonne Paris Cit\'{e} (ANR-10-LABX- 0023 and ANR-11-IDEX-0005-02). Parts of this work were supported by IPGP multidisciplinary PARI program, and by Region \^{I}le-de-France SESAME Grant no. 12015908. M.S. acknowledges funding from the Villum Fonden (no. 00025333). Additional financial support for this project was provided by Carlsberg Foundation (CF18-1105), the Danish National Research Foundation (DNRF97) and the European Research Council (ERC Advanced Grant Agreement, no. 833275-DEEPTIME) to M.B. A.J. was supported by ERC Consolidator Grant 724 687-PLANETESYS.

\appendix

\section{Materials and methods}
\subsection{Samples}
\begin{figure}[ht!]
\centering
\includegraphics[width=1\textwidth]{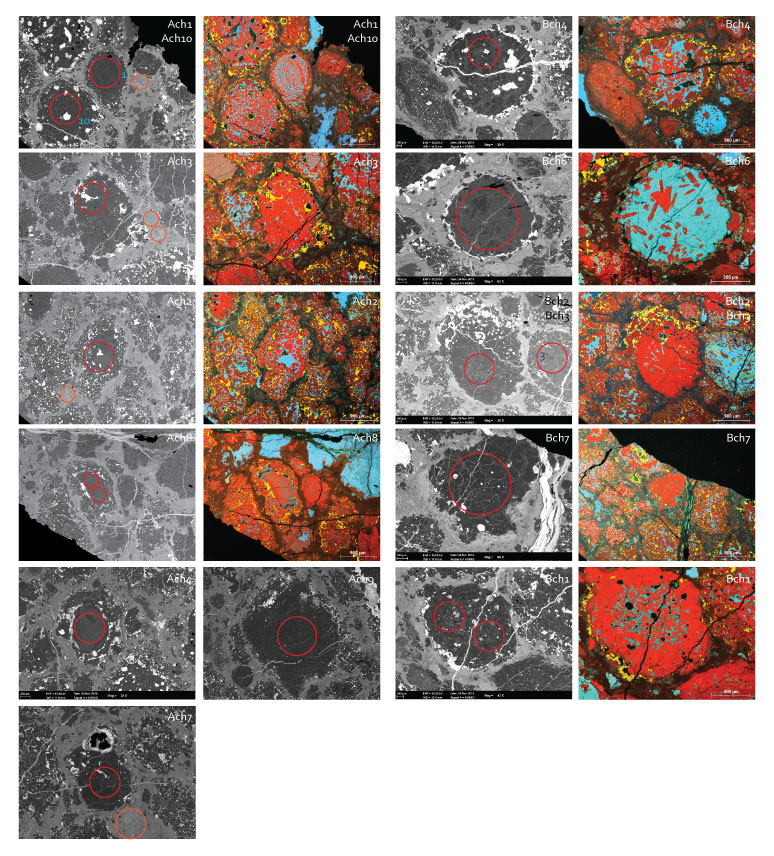}
\caption{Back scattered electron images and Mg-Ca-Al-S elemental maps of chondrules sampled in this study, along with the drill locations (red circles). The orange circles represent matrix locations.\label{fig:petrology}}
\end{figure}

In this study, we have analyzed the Cr and Zn isotope composition, as well as the elemental composition of a total of 37 fractions from the Leoville CV3.1 chondrite, which is a find that belongs to the reduced subgroup of CV chondrites. This CV chondrite was selected based on its low degree of secondary alteration relative to other CV chondrites \citep{bonal_thermal_2016}. Even though the Leoville chondrite experienced some terrestrial weathering in the form of calcium carbonate and Fe-oxide veins throughout the sample \citep{abreu_carbonates_2005}, the boundaries of these veins are sharp, and care was taken to avoid these areas. We sampled a total of 11 matrix areas: 3 areas consist of purely intra-chondrule matrix, 5 samples contain a mixture of fine-grained dust rims and ICM and another 2 samples are sampled from only the FGRs. In addition to these samples, we have measured the Cr isotope composition of an FGR previously analyzed for its Zn isotope composition \citep{van_kooten_zinc_2019}. Likewise, we have analyzed the Cr isotope signatures of 5 chondrule cores (porphyritic olivine type) and 5 corresponding igneous rims, also previously measured for their Zn isotope composition \citep{van_kooten_zinc_2019}. We have sampled an additional 14 chondrules, 11 of which do not contain igneous rims and 2 of them being Al-rich chondrules. We note that part of these samples have been previously analyzed for their composition by laser ablation inductively coupled plasma mass spectrometry \citep{van_kooten_unifying_2019} or by inductively coupled plasma quadrupole mass spectrometry \citep{van_kooten_zinc_2019}. The Leoville fractions were extracted by New Wave micromill with tungsten carbide drill bits at the Institute de Physique du Globe de Paris (IPGP) and thereafter transferred to clean Savillex beakers. The diameter of the drill spots was 150 µm and care was taken not to drill too deep ($<$200 $\upmu$m) to avoid contamination from surrounding materials. This typically resulted in $\sim$100-500 $\upmu$g of material. During drilling, all drillspots were carefully examined under a plain-light microscope and spots where contamination was suspected were discarded. Before and after drilling, back scattered electron (BSE) images of the sampled areas were taken to verify the nature of the sampled materials. This was done using the Zeiss EVO MA10 scanning electron microscope (SEM) at IPGP. In addition to BSE images, we have also made elemental maps of the selected areas (Figs. \ref{fig:petro1} and \ref{fig:petrology}). 

\subsection{Digestion and purification}
All extracted samples were first digested in concentrated mixtures of 1:1 HNO$_{3}$/HF acids using PARR bombs at temperatures of 210 $^{\circ}$C for 48 hours, to ensure complete digestion of potential refractory minerals located in the Leoville fractions. The samples were then dried down and taken up in aqua regia to dissolve fluorides formed during the first digestion step. The Leoville fractions were taken up in 200 $\upmu$l 1.5M HBr and 5 \% of the aliquot was saved for elemental composition analyses by Agilent 7900 ICP-Q-MS at IPGP. The samples were first eluted on a 100 $\upmu$l AG1X8 (200–400 mesh) anion resin using 1.5M HBr to separate all elements from Zn, which was subsequently eluted in 0.5M HNO$_{3}$ \citep{van_kooten_zinc_2019}. This purification step was repeated to ensure the complete removal of all impurities from the Zn cut. The final Zn cut was between 5-15 ng for all Leoville fractions and the procedural blank was $<$0.2 ng. The first HBr cut eluted from the anion resin was saved for further Cr purification. The second HBr cut of the repeat column was not added to the first cut, since it contains a negligible amount of Cr (0.3 wt.\% of the total). These aliquots were dried down and taken up in 230 $\upmu$l 2M HNO$_{3}$ and 20 $\upmu$l concentrated H2O2. The samples were left at room temperature for $>$1 week to ensure complete speciation of Cr to Cr$^{3+}$. Similar to the Zn purification method \citep{van_kooten_zinc_2019}, an adjusted Cr column chromatography was developed to accommodate the small sample sizes of $\sim$100 $\upmu$g ($\sim$30 ng Cr). This involves a two-in-one column procedure that is fast (3 hours cleaning, 3 hours elution), produces low blanks ($<$0.2 ng Cr) and avoids the risk of losing these small samples by repeated dry-down and take-up of acids. In detail, two columns are stacked on top of each other, with a TOGDA column (250 $\upmu$l resin volume) on top of a biorad AG50X8 (200–400 mesh) cation (1 ml resin volume). The samples are loaded on the top column in their pretreatment solution of 2M HNO$_{3}$/H$_{2}$O$_{2}$ and are subsequently eluted with 1 ml of 2M HNO$_{3}$, thereby efficiently retaining Ca on the resin. This elute drops directly onto a preconditioned cation column and, after removing the top column, an additional 750 $\upmu$l of 2M HNO$_{3}$ are added. The total elution of 2 ml 2M HNO$_{3}$ on the cation ensures the separation of Na and K. We note that the separation of Cr and matrix elements is also achieved by eluting the latter with a 2M HNO$_{3}$/HF solution. However, we describe the procedure here to include the further separation of Na$\pm$K (2 ml 2M HNO$_{3}$), Fe|Ti|V|Al (5 ml 1M HF) and Mg|Ni|Mn (22 ml 1M HNO$_{3}$) for future analyses. The final elution of Cr is preceded by a last clean-up of Fe, Ti and V in 3 ml 1M HF (to avoid interferences on the Cr signal by TIMS) and is done using 10 ml of 6M HCl. The total recovery of Cr is $>$95 \%, which has been tested using various rock compositions, including BHVO-2, PCC1, bulk CV and CI chondrites. After chemistry, the dried Zn- and Cr-cuts are finally taken up in 7M HNO$_{3}$ and are heated to 140 $^{\circ}$C for $>$24 hours before isotope analyses.

\subsection{Neptune Plus MC-ICPMS and Triton TIMS}
\begin{figure}[ht!]
\centering
\includegraphics[width=0.8\textwidth]{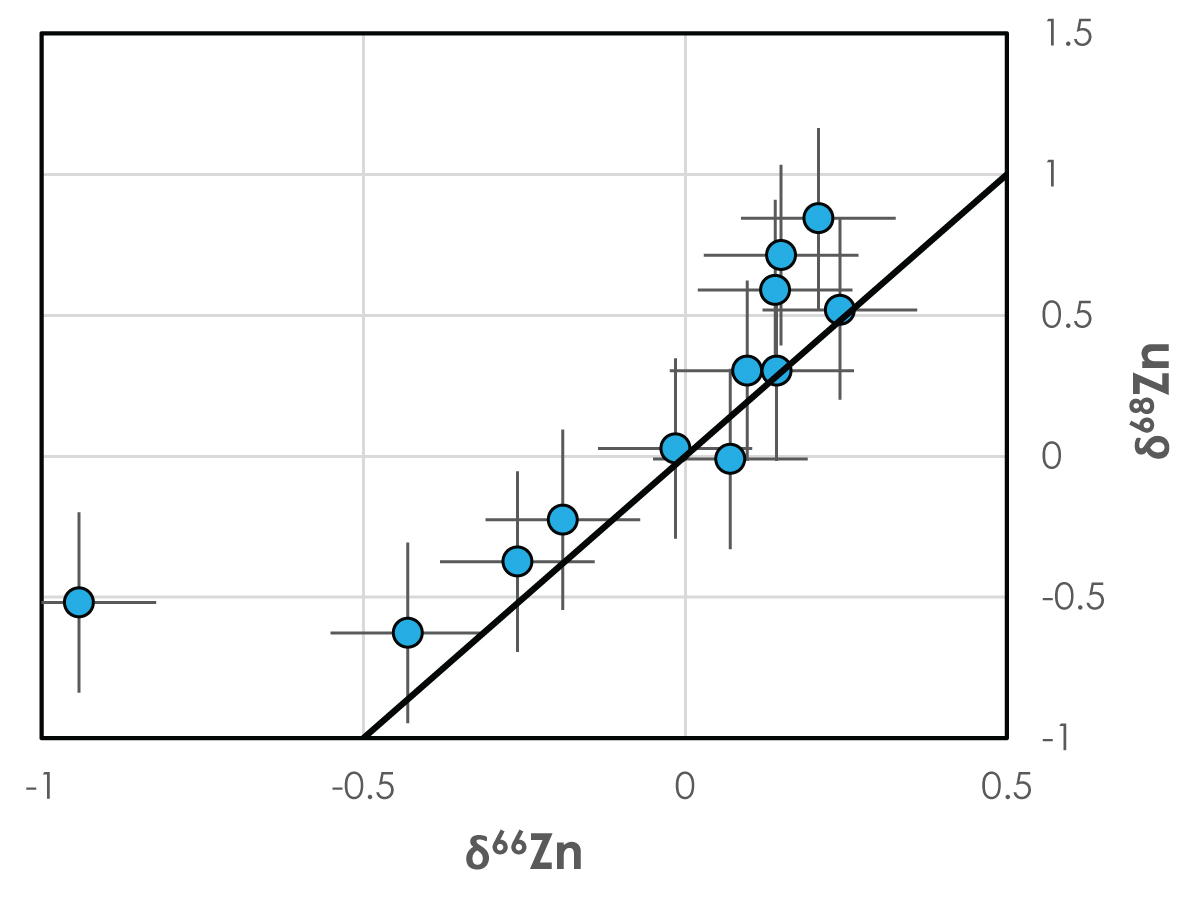}
\caption{Zn isotope analyses of Leoville matrix and chondrule fractions, showing $\updelta$$^{66}$Zn versus $\updelta$$^{68}$Zn values. The solid line reflects the mass-dependent relationship between both isotope ratios. \label{fig:massdependent}}
\end{figure}

Zinc isotopes ($^{64}$Zn, $^{66}$Zn, $^{68}$Zn) were measured using a Thermo Scientific Neptune Plus Multi-Collector Inductively-Coupled-Plasma Mass-Spectrometer (MC-ICPMS) at IPGP and the analytical setup was done according to \citet{van_kooten_zinc_2019}. BHVO-2 and CV bulk chondrite standards were measured alongside the samples to provide an estimate of the accuracy and reproducibility of the analyses within each session. The samples were bracketed by the JMC-Lyon standard and the data are reported in the delta notation as permil deviations from the standard ($\updelta$$^{66}$Zn = [[$^{66}$Zn/$^{64}$Zn$_{smp}$]/[$^{66}$Zn/$^{64}$Zn$_{std}$]-1] $\times$ 1000). The mass-dependent relationship between the $\updelta$$^{66}$Zn and $\updelta$$^{68}$Zn data are shown in Fig. \ref{fig:massdependent}.\\

The Cr-isotope composition of all samples was measured by thermal ionization mass spectrometry (Triton TIMS) at the Centre for Star and Planet Formation (StarPlan, Denmark). Chromium isotope analyses were conducted using a hybrid method of total evaporation and standard-sample bracketing technique fully described in \citet{van_kooten_isotopic_2016}. Samples typically contained 30–100 ng Cr and were measured on W-filaments with 15–30 ng per filament. Considering potential analytical biases related to the processing of such small sample sizes, we have thoroughly tested this method using different rock standards that encompass the chemical compositions of the various Leoville fractions. This included the Cr purification of five individual 100 $\upmu$g (total sample) aliquots of the CI chondrite Ivuna (representing the matrix), which were measured on five individual filaments and yield an average $\upepsilon$$^{54}$Cr value of 1.50$\pm$0.23 ‱ (the epsilon notation reflects the parts per thousand deviation of the mass bias corrected $^{x}$Cr/$^{52}$Cr ratio from the SRM 979 standard, where x = 53 or 54). We also processed three aliquots of 100 $\upmu$g each from the bulk CV chondrite NWA 12523 as well as two aliquots from the PCC1 geological standard (serving as analogues for bulk chondrules), which yield average $\upepsilon$$^{54}$Cr values of 0.81$\pm$0.28 and 0.12$\pm$0.09 ‱, respectively. Hence, we show that such small fractions ($\sim$30 ng Cr) can be successfully measured with an external reproducibility of $<$30 ppm (2SD). Finally, as a last test to verify our TIMS results, we have put another, larger ($\sim$2 $\upmu$g Cr) aliquot of same digestion used for TIMS analyses of the CV chondrite NWA 12523 through the same chemistry and measured the Cr isotope composition by MC-ICPMS (StarPlan). In addition, we were able to run the largest chondrule sample LeoBch4 by TIMS as well as MC-ICPMS. The MC-ICPMS analyses yield $\upepsilon$$^{54}$Cr values that are indistinguishable from the TIMS data (CV: 0.92±0.12 ‱, LeoBch4: 0.26$\pm$0.30 ‱ [TIMS: 0.18$\pm$0.30 ‱ ]).

\section{Petrology and compositions of Leoville fractions}
We have carefully examined the petrological fabric of all fractions before subjecting them to microdrill extraction and have, thus, managed to sample a wide range of compositions and textures that together reflect the entirety of material accreted to form the CV chondrites (Fig. \ref{fig:petro1}). Firstly, the chondrules of Leoville are typically between 1-3 mm in diameter and consist dominantly of porphyritic olivine textures. Approximately 10 vol.\% of the chondrules have barred olivine textures, 50 vol.\% contain relatively fine-grained olivine-pyroxene igneous rims with abundant metal and sulfide and about 30 vol.\% contain mineralogical zonation of increasing pyroxene or olivine towards the chondrule core rim relative to the feldspathic mesostasis. Most chondrules contain one or more metal/sulfide rims around their primary or secondary cores and the surface area of these rims often surpasses that of the cores (Fig. \ref{fig:petro1}, Fig. \ref{fig:petrology}). About 5 vol.\% of all chondrules in our thick section of Leoville are Al-rich (i.e., $>$10 wt.\% of Al$_{2}$O$_{3}$). Secondly, all chondrules are surrounded by fine-grained dust rims (FGRs), which are typically between 100 and 300 $\upmu$m in diameter. The space between the FGRs is occupied by a more coarse-grained matrix (ICM: intra-chondrule matrix), which includes chondrule fragments and larger sulfide and metal grains. 

We provide the major and minor element compositions of Leoville chondrules, rims and matrices in Table A1. Since the compositions are derived from solutions obtained through HF acid digestion, the exclusion of Si in these measurements prevented element normalization and, hence, we discuss only element ratios. Foremost, these ratios can be used to compare laser ablation (LA) ICPMS derived compositions of Leoville fractions \citep{van_kooten_unifying_2019} to microdrill extracted aliquots and potential contaminations related to this technique. For example, the sampling of various types of matrix presents great difficulties due to the small size of the areas relative to the drill tip. While LA analyses of the FGRs show this matter to be CI-like, these drilled areas may include contaminations from surrounding chondrules and/or CAIs. As a result, of the ten attempts to sample these FGRs, only one proved successful. The matrix composition of Ch2 is the only sample with a near-solar Al/Mg ratio of 0.12 and a super-chondritic Fe/Mg ratio of 3.4, which is indicative of pristine CV chondrite FGRs \citep{van_kooten_unifying_2019}. Other matrix samples have either higher Al/Mg ratios (indicative of CAI contamination), and/or lower Fe/Mg ratios (from chondrule contamination). These fractions have either been designated ‘contaminated’ or ‘clean’ ICM, depending on the visual inspection of the drill holes. Two of the sampled chondrules are Al-rich chondrules, based on their high Al/Mg, Ca/Mg and Na/Mg ratios. 

\section{Mass balance calculations of CV chondrite components}
We outline here the mass balance of relatively unaltered CV chondritic components using the Cr isotope data from Leoville chondrules, igneous rims, Al-rich chondrules, fine-grained dust rims, intra-chondrule matrix and CAIs (Table A3). We note that this mass balance is subject to the errors of the Cr isotope measurements and the estimations of the modal abundances of the components. We have calculated the bulk CV $\upepsilon$$^{54}$Cr value in a scenario where all matrix, including FGRs, have a $\upepsilon$$^{54}$Cr composition similar to the average ICM ($\upepsilon$$^{54}$Cr = 0.7 ‱). We note that although the values of Al-rich chondrules may be widely variable (Table \ref{tab:CrZn}), possibly due to incorporation of different CAI- and FUN-CAI-like precursors, their total budget is estimated to be $<$1 vol.\% of all chondrules \citep{zhang_origin_2019} and is thus insignificant for the calculated bulk chondrule $\upepsilon$$^{54}$Cr value, which is around --0.5 ‱. The bulk CV chondrite value including only chondrules and matrix is then estimated to be $\sim$0 ‱, assuming both components have similar Cr contents ($\sim$3000 ppm). CAIs and AOAs have much lower Cr contents ($\sim$200 ppm) and although their$\upepsilon$$^{54}$Cr values are high (6.8 and 5.4 ‱, respectively; \citealp{larsen_evidence_2011}), the mass balance of chondrules, matrix and CAIs shows that the final calculated bulk $\upepsilon$$^{54}$Cr value is $<$0.1 ‱ (Table A3). It is currently unknown what the contribution of the metal (not related to the chondrules) to the Cr isotope budget is. Leaching experiments of Leoville show that the metal may have a negative $\upepsilon$$^{54}$Cr signature \citep{trinquier_widespread_2007}. Hence, the mass balance of all CV chondrite components suggests that the CI-like $\upepsilon$$^{54}$Cr value measured for the FGR in this study is likely applicable to all FGRs and is required to get a bulk CV chondrite value of 0.8 ‱.

\begin{table}[]
\caption{Mass balance calculations of CV chondritic components and their $\upepsilon$$^{54}$Cr values. The bulk CV $\upepsilon$$^{54}$Cr value ($<$0.1) is calculated as if all matrix had the composition of the average ICM. The measured CV bulk value (0.8) is the measured $\upepsilon$$^{54}$Cr value. \label{fig:massbalance}}
\begin{tabular}{llllll}
\toprule
\textbf{Component} & \multicolumn{2}{l}{\textbf{Modal abundance (vol.\%)}} & \multicolumn{3}{l}{\textbf{$\upepsilon$$^{54}$Cr}} \\
\midrule
Chondrule core & 75 & \multirow{3}{*}{\} 45$^{a}$} & -0.6 & \multirow{3}{*}{\} -0.5} & \multirow{5}{*}{\} 0} \\
Chondrule rim & 25 &  & -0.1 &  &  \\
Al-rich chondrules & \textless{}1$^{c}$ &  & $\sim$0? &  &  \\
\midrule
FGR & 50 & \multirow{2}{*}{\} 40$^{a}$} & 1.3 & \multirow{2}{*}{\}   0.7$^{d}$} &  \\
ICM & 50 &  & 0.7 &  &  \\
\midrule
CAIs/AOAs &  & 10$^{a}$ & 6.8 &  & 6.8 \\
metal$^{b}$ &  & \textless{}5 & ? &  &  \\
\midrule
Bulk &  & 100 & 0.8 &  & \textless{}0.1
\end{tabular}
\end{table}

\section{Implications from potential $^{26}$Al heterogeneity in the protoplanetary disk}
Variations in $\upmu^{26}$Mg*  (the decay product of $^{26}$Al) between bulk Solar System materials and individual components such as chondrules have been interpreted to reflect $^{26}$Al \citep{bollard_early_2017,larsen_evidence_2011,olsen_magnesium_2016,van_kooten_isotopic_2016,van_kooten_magnesium_2017,van_kooten_role_2020,schiller_early_2015,connelly_absolute_2018,connelly_pbpb_2017,larsen_accretion_2016} and Mg isotope heterogeneity \citep{budde_hf-w_2018,wasserburg_mg_2012}. Typically, the accretion timescales of chondrites, achondrites and their parent bodies are modelled assuming a homogeneous and canonical distribution of $^{26}$Al. However, if the interpretation of a reduced $^{26}$Al reservoir for chondrites is correct \citep{larsen_evidence_2011,schiller_early_2015}, then accretion of chondrites was more rapid than assuming a canonical $^{26}$Al/$^{27}$Al ratio.


\begin{sidewaystable}[]
\centering
\caption{Elemental compositions and ratios of Leoville drilled fractions by ICP-Q-MS (see Appendix A for details). Concentrations are given in ppb together with the relative standard deviation (RSD\%).} 
\tiny
\begin{tabular}{lrlrlrlrlrlrlrlrlrlrlrlrlrl}
& \multicolumn{10}{c}{\textit{Cores}} & \multicolumn{14}{c}{\textit{ICM}} & \multicolumn{2}{c}{\textit{FGR}} \\
\cmidrule{3-10}
\cmidrule{13-24}
 & \multicolumn{2}{c}{\textbf{Ach3}} & \multicolumn{2}{c}{\textbf{Ach1}} & \multicolumn{2}{c}{\textbf{Bch6}} & \multicolumn{2}{c}{\textbf{Ach7}} & \multicolumn{2}{c}{\textbf{Ach2}} & \multicolumn{2}{c}{\textbf{Bmx1}} & \multicolumn{2}{c}{\textbf{Bmx2}} & \multicolumn{2}{c}{\textbf{Bmx3}} & \multicolumn{2}{c}{\textbf{Ach7mx}} & \multicolumn{2}{c}{\textbf{Ach1mx}} & \multicolumn{2}{c}{\textbf{Ach2mx}} & \multicolumn{2}{c}{\textbf{Ach3mx}} & \multicolumn{2}{c}{\textbf{Ch2mx}} \\
 &  &  &  &  &  &  &  &  &  &  &  &  &  &  &  &  &  &  &  &  &  &  &  &  &  &  \\
Li & 0.02 & $\pm$42 & 0.04 & $\pm$23 & 0.06 & $\pm$18 & 0.04 & $\pm$17 & 0.04 & $\pm$27 & 0.04 & $\pm$19 & 0.04 & $\pm$12 & 0.05 & $\pm$9 & 0.05 & $\pm$4 & 0.05 & $\pm$8 & 0.05 & $\pm$9 & 0.04 & $\pm$9 & 0.020 & $\pm$4 \\
Na & 4.3 & $\pm$3 & 6.2 & $\pm$3 & 15.1 & $\pm$1 & 2.2 & $\pm$5 & 8.6 & $\pm$1 & 2.9 & $\pm$4 & 4.4 & $\pm$7 & 3.9 & $\pm$8 & 16.9 & $\pm$2 & 3.0 & $\pm$11 & 5.5 & $\pm$6 & 4.2 & $\pm$7 & 3.7 & $\pm$4 \\
Mg & 1097 & $\pm$0.4 & 484 & $\pm$0.1 & 91 & $\pm$0.4 & 610 & $\pm$0.3 & 708 & $\pm$0.1 & 90 & $\pm$1.1 & 138 & $\pm$0.2 & 131 & $\pm$0.2 & 321 & $\pm$0.9 & 124 & $\pm$0.2 & 148 & $\pm$0.8 & 82 & $\pm$0.3 & 83.6 & $\pm$2 \\
Al & 34.7 & $\pm$2 & 71.7 & $\pm$1 & 58.6 & $\pm$2 & 11.0 & $\pm$1 & 62.0 & $\pm$1 & 7.9 & $\pm$4 & 11.4 & $\pm$2 & 15.2 & $\pm$2 & 31.7 & $\pm$2 & 12.7 & $\pm$3 & 15.5 & $\pm$4 & 9.9 & $\pm$3 & 8.3 & $\pm$2 \\
P & 4.1 & $\pm$28 & 4.6 & $\pm$32 & 18.5 & $\pm$11 & 5.2 & $\pm$53 & 4.1 & $\pm$70 & 5.5 & $\pm$43 & 6.7 & $\pm$20 & 2.2 & $\pm$68 & 4.0 & $\pm$28 & 7.5 & $\pm$25 & \multicolumn{2}{c}{n.d.} & 5.9 & $\pm$45 & \multicolumn{2}{c}{n.d.} \\
K & 4.5 & $\pm$4 & 3.8 & $\pm$6 & 6.1 & $\pm$9 & 3.5 & $\pm$12 & 3.6 & $\pm$6 & 3.2 & $\pm$15 & 4.3 & $\pm$46 & 4.1 & $\pm$49 & 2.4 & $\pm$91 & 2.3 & $\pm$84 & 1.6 & $\pm$127 & \multicolumn{2}{c}{n.d.} & \multicolumn{2}{c}{n.d.} \\
Ca & 32.9 & $\pm$7 & 68.9 & $\pm$3 & 43.8 & $\pm$7 & 14.3 & $\pm$3 & 64.1 & $\pm$4 & 9.6 & $\pm$17 & 13.3 & $\pm$8 & 15.5 & $\pm$12 & 33.9 & $\pm$3 & 12.4 & $\pm$11 & 17.6 & $\pm$6 & 12.1 & $\pm$5 & 8.4 & $\pm$7 \\
Ti & 2.3 & $\pm$6 & 3.4 & $\pm$4 & 1.6 & $\pm$13 & 0.7 & $\pm$6 & 3.1 & $\pm$4 & 0.3 & $\pm$13 & 0.6 & $\pm$13 & 0.8 & $\pm$13 & 1.7 & $\pm$6 & 0.5 & $\pm$22 & 0.8 & $\pm$3 & 0.4 & $\pm$24 & 0.3 & $\pm$31 \\
V & 0.38 & $\pm$2 & 0.23 & $\pm$4 & 0.56 & $\pm$1 & 0.19 & $\pm$3 & 0.34 & $\pm$2 & 0.05 & $\pm$6 & 0.07 & $\pm$8 & 0.07 & $\pm$2 & 0.18 & $\pm$0.5 & 0.07 & $\pm$2 & 0.08 & $\pm$3 & 0.04 & $\pm$8 & 0.05 & $\pm$0.5 \\
Cr & 7.2 & $\pm$0.7 & 5.0 & $\pm$0.3 & 6.0 & $\pm$0.5 & 5.1 & $\pm$0.7 & 11.7 & $\pm$0.6 & 2.5 & $\pm$1.2 & 3.8 & $\pm$0.6 & 3.4 & $\pm$1.2 & 8.3 & $\pm$0.1 & 3.5 & $\pm$1.0 & 4.2 & $\pm$1.2 & 2.1 & $\pm$1.4 & 2.5 & $\pm$2 \\
Mn & 2.0 & $\pm$0.9 & 1.3 & $\pm$1.3 & 1.1 & $\pm$2.4 & 0.8 & $\pm$2.4 & 1.8 & $\pm$1.0 & 1.3 & $\pm$1.1 & 1.9 & $\pm$2.3 & 2.1 & $\pm$1.1 & 5.0 & $\pm$0.5 & 1.8 & $\pm$0.8 & 2.1 & $\pm$2.9 & 1.8 & $\pm$0.6 & 1.6 & $\pm$2 \\
Fe & 169 & $\pm$0.6 & 79 & $\pm$0.4 & 163 & $\pm$0.3 & 304 & $\pm$0.2 & 584 & $\pm$0.2 & 204 & $\pm$0.6 & 262 & $\pm$0.1 & 274 & $\pm$0.5 & 590 & $\pm$0.4 & 260 & $\pm$0.3 & 249 & $\pm$0.7 & 241 & $\pm$0.2 & 280.0 & $\pm$1 \\
Co & 0.8 & $\pm$1.5 & 0.4 & $\pm$1.9 & 0.8 & $\pm$0.6 & 0.6 & $\pm$1.0 & 2.4 & $\pm$1.4 & 0.6 & $\pm$1.1 & 1.0 & $\pm$1.4 & 0.6 & $\pm$2.8 & 0.9 & $\pm$0.9 & 0.6 & $\pm$1.4 & 0.9 & $\pm$1.5 & 0.5 & $\pm$1.1 & 0.6 & $\pm$0.5 \\
Ni & 8.3 & $\pm$0.4 & 1.9 & $\pm$1.6 & 20.0 & $\pm$0.7 & 3.9 & $\pm$0.5 & 48.3 & $\pm$0.3 & 11.5 & $\pm$0.7 & 15.5 & $\pm$0.9 & 19.2 & $\pm$0.7 & 14.2 & $\pm$1.4 & 13.9 & $\pm$1.0 & 15.6 & $\pm$1.3 & 9.2 & $\pm$1.3 & 10.9 & $\pm$1 \\
Cu & 0.09 & $\pm$3 & 0.31 & $\pm$1 & 0.24 & $\pm$2 & 0.05 & $\pm$4 & 0.35 & $\pm$1 & 0.11 & $\pm$3 & 0.16 & $\pm$4 & 0.15 & $\pm$1 & 0.15 & $\pm$2 & 0.13 & $\pm$3 & 0.15 & $\pm$1 & 0.09 & $\pm$4 & \multicolumn{2}{c}{n.d.} \\
Zn & 0.04 & $\pm$28 & 0.19 & $\pm$16 & 0.42 & $\pm$3 & 0.01 & $\pm$49 & 0.09 & $\pm$11 & 0.07 & $\pm$34 & 0.13 & $\pm$9 & 0.13 & $\pm$13 & 0.25 & $\pm$7 & 0.13 & $\pm$7 & 0.27 & $\pm$4 & 0.11 & $\pm$3 & 0.29 & $\pm$5 \\
Rb & 0.016 & $\pm$47 & 0.014 & $\pm$54 & 0.019 & $\pm$37 & 0.013 & $\pm$57 & \multicolumn{2}{c}{n.d.} & \multicolumn{2}{c}{n.d.} & 0.016 & $\pm$23 & 0.017 & $\pm$22 & 0.013 & $\pm$30 & 0.015 & $\pm$24 & 0.017 & $\pm$22 & 0.010 & $\pm$37 & \multicolumn{2}{c}{n.d.} \\
Sr & 0.024 & $\pm$2 & 0.059 & $\pm$3 & 0.059 & $\pm$2 & 0.014 & $\pm$4 & 0.064 & $\pm$2 & 0.005 & $\pm$9 & 0.008 & $\pm$7 & 0.032 & $\pm$3 & 0.026 & $\pm$4 & 0.013 & $\pm$4 & 0.014 & $\pm$4 & 0.008 & $\pm$7 & 0.008 & $\pm$5 \\
Sn & 0.23 & $\pm$12 & 0.09 & $\pm$33 & 0.24 & $\pm$20 & 0.12 & $\pm$21 & 0.08 & $\pm$30 & 0.15 & $\pm$20 & 0.12 & $\pm$12 & 0.11 & $\pm$15 & 0.10 & $\pm$15 & 0.09 & $\pm$14 & 0.08 & $\pm$16 & 0.05 & $\pm$19 &  &  \\
Sb & 0.19 & $\pm$4 & 0.15 & $\pm$5 & 0.23 & $\pm$3 & 0.17 & $\pm$5 & 0.14 & $\pm$5 & 0.18 & $\pm$4 & 0.19 & $\pm$0.5 & 0.18 & $\pm$3 & 0.17 & $\pm$2 & 0.16 & $\pm$3 & 0.14 & $\pm$0.5 & 0.13 & $\pm$4 &  &  \\
Ba & \multicolumn{2}{c}{n.d.} & 0.02 & $\pm$4 & 0.06 & $\pm$14 & \multicolumn{2}{c}{n.d.} & 0.01 & $\pm$10 & \multicolumn{2}{c}{n.d.} & 0.01 & $\pm$49 & 0.15 & $\pm$4 & 0.01 & $\pm$25 & 0.10 & $\pm$4 & 0.01 & $\pm$48 & 0.01 & $\pm$34 & 0.01 & $\pm$10 \\
 &  &  &  &  &  &  &  &  &  &  &  &  &  &  &  &  &  &  &  &  &  &  &  &  &  &  \\
Fe/Mg & 0.15 & $\pm$0.7 & 0.16 & $\pm$0.4 & 1.79 & $\pm$0.5 & 0.50 & $\pm$0.4 & 0.83 & $\pm$0.3 & 2.26 & $\pm$1.3 & 1.90 & $\pm$0.2 & 2.09 & $\pm$0.5 & 1.84 & $\pm$0.9 & 2.09 & $\pm$0.3 & 1.68 & $\pm$1.0 & 2.94 & $\pm$0.3 & 3.35 & $\pm$1.9 \\
$^{27}$Al/$^{24}$Mg & 0.04 & $\pm$2 & 0.19 & $\pm$1 & 0.81 & $\pm$2 & 0.02 & $\pm$1 & 0.11 & $\pm$1 & 0.11 & $\pm$4 & 0.10 & $\pm$2 & 0.14 & $\pm$2 & 0.12 & $\pm$2 & 0.13 & $\pm$3 & 0.13 & $\pm$4 & 0.15 & $\pm$3 & 0.12 & $\pm$3 \\
$^{55}$Mn/$^{52}$Cr & 0.33 & $\pm$1 & 0.31 & $\pm$1 & 0.21 & $\pm$2 & 0.19 & $\pm$2 & 0.18 & $\pm$1 & 0.64 & $\pm$2 & 0.58 & $\pm$2 & 0.74 & $\pm$2 & 0.72 & $\pm$1 & 0.62 & $\pm$1 & 0.60 & $\pm$3 & 1.00 & $\pm$2 & 0.51 & $\pm$2 \\
\end{tabular}
\end{sidewaystable}

\begin{table}[]
\centering
\caption{Overview table of analyses carried out for samples used in this study. BSE = back scattered electron images, maps = elemental maps from SEM, LA = laser ablation ICPMS and WA = wet analyses by ICP-Q-MS or HR-ICPMS (see Appendix A for detailed explanation). Analyses were done in this study [*], by \citet{van_kooten_unifying_2019} [1] or by \citet{van_kooten_zinc_2019} [2].}
\begin{tabular}{lccccc}
             & \textbf{BSE} & \textbf{maps} & \textbf{LA} & \textbf{WA} & \textbf{ref}  \\
             \midrule
\textit{Cores }        &     &      &    &     &         \\
Ach1          & x   & x    &    &     & {[}*{]} \\
Ach7          & x   &      &    &     & {[}*{]} \\
Bch6          & x   & x    & x  &     & {[}1{]} \\
C1            & x   &      &    & x   & {[}2{]} \\
C2            & x   &      &    & x   & {[}2{]} \\
C3            & x   &      &    & x   & {[}2{]} \\
C5            & x   &      &    & x   & {[}2{]} \\
C6            & x   &      &    & x   & {[}2{]} \\
Bch1          & x   & x    &    & x   & {[}*{]} \\
Bch2          & x   & x    &    & x   & {[}*{]} \\
Bch3          & x   & x    &    & x   & {[}*{]} \\
Bch4          & x   & x    &    & x   & {[}*{]} \\
Bch7          & x   & x    &    & x   & {[}*{]} \\
Ach4          & x   &      &    & x   & {[}*{]} \\
Ach8          & x   & x    &    & x   & {[}*{]} \\
Ach9          & x   &      &    & x   & {[}*{]} \\
Ach10         & x   & x    &    & x   & {[}*{]} \\
              &     &      &    &     &         \\
\textit{Core with rim} &     &      &    &     &         \\
Ach2          & x   & x    & x  &     & {[}1{]} \\
Ach3          & x   & x    &    &     & {[}*{]} \\
              &     &      &    &     &         \\
\textit{Rims}          &     &      &    &     &         \\
C1            & x   & x    &    & x   & {[}2{]} \\
C2            & x   & x    &    & x   & {[}2{]} \\
C3            & x   & x    &    & x   & {[}2{]} \\
C5            & x   & x    &    & x   & {[}2{]} \\
C6            & x   & x    &    & x   & {[}2{]} \\
              &     &      &    &     &         \\
\textit{ICM}           &     &      &    &     &         \\
Ach1mx        & x   & x    &    & x   & {[}*{]} \\
Ach2mx        & x   & x    &    & x   & {[}*{]} \\
Ach3mx        & x   & x    &    & x   & {[}*{]} \\
Ach7mx        & x   &      &    & x   & {[}*{]} \\
Bmx1          & x   &      &    & x   & {[}*{]} \\
Bmx2          & x   &      &    & x   & {[}*{]} \\
Bmx3          & x   &      &    & x   & {[}*{]} \\
Leo2ch1mx1    & x   &      &    &     & {[}*{]} \\
Leo2ch1mx2    & x   &      &    &     & {[}*{]} \\
Leo2ch2mx     & x   &      &    &     & {[}*{]} \\
Leo2ch3mx     & x   &      &    &     & {[}*{]} \\
              &     &      &    &     &         \\
\textit{FGR}          &     &      &    &     &         \\
C1            & x   & x    &    & x   & {[}2{]} \\
C2            & x   & x    &    & x   & {[}2{]}
\end{tabular}
\end{table}

\end{document}